\documentclass[a4paper,fleqn,usenatbib,useAMS]{mnras}

\usepackage{graphicx}	
\usepackage{amsmath}	
\usepackage{amssymb}	
\usepackage{multicol}        
\usepackage{bm}		
\usepackage{pdflscape}	
\usepackage{longtable}	
\usepackage{soul}

\usepackage[T1]{fontenc}
\usepackage{ae,aecompl}

\usepackage{newtxtext,newtxmath}

\title[Probing Saraswati's heart]{Probing Saraswati's heart: evaluating the dynamical state of the massive galaxy cluster A2631 through a comprehensive weak-lensing and dynamical analysis}

\author[Monteiro-Oliveira et al.] 
  {R.~Monteiro-Oliveira,$^{1,2}$\thanks{E-mail: rogerionline@gmail.com}
  A. C.~Soja,$^3$
  A. L. B.~Ribeiro,$^2$
  J.~Bagchi,$^4$  
  S.~Sankhyayan,$^5$
  \newauthor 
  T. O.~Candido,$^3$
  and
  R. R.~Flores$^3$\\
  $^1$Universidade de S\~ao Paulo, Inst. de Astronomia, Geof\'isica e Ci\^encias Atmosf\'ericas, Depto. de Astronomia, R. do Mat\~ao 1226, 05508-090 S\~ao Paulo, Brazil\\
  $^2$Laborat\'orio de Astrof\'isica Te\'orica e Observacional,  Depto. de Ci\^encias Exatas e Tecnol\'ogicas,  Universidade Estadual de Santa Cruz, 45650-000, Ilh\'eus, BA, Brazil\\  
  $^3$Instituto Federal do Mato Grosso do Sul, Campus Corumb\'a, Rua Pedro de Medeiros s/n, 79310-110 Corumb\'a, Brazil\\  
  $^4$Inter-University Centre for Astronomy and Astrophysics, Post Bag 4, Ganeshkhind, Pune 411007, India\\ 
  $^5$National Centre for Radio Astrophysics, TIFR, Post Bag 3, Ganeshkhind, Pune - 411007, India\\ 
 }

\date{Accepted 2020 November 13. Received 2020 November 13; in original form 2020 July 06}

\pubyear{2018}

\begin{document}
\label{firstpage}
\pagerange{\pageref{firstpage}--\pageref{lastpage}}
\maketitle

\begin{abstract}
In this work, we investigate the dynamical state of the galaxy cluster Abell 2631, a massive structure located at the core of the Saraswati supercluster. To do this, we first solve a tension found in the literature regarding the weak-lensing mass determination of the cluster. We do this through a comprehensive weak-lensing analysis, exploring the power of the combination of shear and magnification data sets. We find  $M_{200}^{\rm wl} = 8.7_{-2.9}^{+2.5} \times 10^{14}$ M$_\odot$. We also determined the mass based on the dynamics of spectroscopic members, corresponding to  $M_{200}^{\rm dy} = 12.2\pm3.0 \times 10^{14}$ M$_\odot$, consistent within a 68 per cent CL with the weak-lensing estimate. The scenarios provided by the mass distribution and dynamics of galaxies are reconciled with those provided by X-ray observations in a scenario where A2631 is observed at a late stage of merging. 
\end{abstract}

\begin{keywords}
gravitational lensing: weak -- dark matter --  clusters: individual: Abell~2631 -- large-scale structure of Universe
\end{keywords}



\section{Introduction}
\label{sec:intro}

At the largest scales, the matter distribution of the Universe matches a web-like structure surrounded by voids where the density is particularly low \citep[e.g.][]{Springel18}. This picture has been designed across time by the gravitational interaction within the context of the hierarchical scenario, where the largest structures are formed late through the merger of the smallest. In this context, superclusters of galaxies constitute the next generation of the most massive large-scale structures in the Universe \citep[e.g.][]{Lacey93, Kravtsov12}.

The supercluster Saraswati was discovered by \cite{Bagchi17} in the Stripe 82 region of the Sloan Digital Sky Survey \citep[SDSS;][]{York00}. It forms a wall-like structure covering $\sim$200 Mpc at $z\sim0.28$. The main body of the supercluster comprises at least 43 galaxy clusters or groups  with a total mass of $\sim2 \times 10^{16} M_{\odot}$, which includes  at least 23  massive galaxy clusters of $M_{200}> 1\times 10^{14} M_{\odot}$ according to the mass-richness relation. This implies a peculiar high-mass concentration since only $\approx 2$ massive galaxy clusters are expected within  Saraswati's whole volume according to the excursion set approach \citep{Sheth01}. The bound core  ($r\sim 20$ Mpc) is composed of five high-mass galaxy clusters. These properties place Saraswati among the few largest and most massive superclusters known, comparable to the most massive `Shapley Concentration' or `Shapley Attractor’ ($z=0.046$) in the nearby universe \citep{Melnick87,Scaramella89,Raychaudhury89}.

Due to a whole range of matter overdensities present in the cosmic web, ranging from rarified voids, intermediate-mass filaments and groups to highly overdense massive galaxy clusters, this supercluster offers exciting possibilities for several promising studies, and one such method is the gravitational weak-lensing mapping of dark matter distributed in the vast supercluster environment, which has  rarely been attempted. Such studies provide a good understanding of what physical processes were involved in the growth of such enormous cosmic structures in the distant universe ($\sim 4$ Gyr lookback time) when mysterious dark energy had just started to dominate structure formation.

At the very centre of the bound core, or the `heart' of Saraswati, is located the most massive cluster member, Abell 2631 ($z = 0.277$), which is an extremely rich (Abell richness class $R = 3$), massive ($ M_{500} \simeq 10^{15} M_{\odot}$) and hot ($T_{e} \simeq 8$ KeV) galaxy cluster \citep{Bagchi17}. This cluster is the subject matter  of our present paper.

Abell 2631, also known as RXCJ~2337.6+0016 (hereafter A2631), has been studied through several wavelengths. For example, using XMM-{\it Newton} data, \cite{Finoguenov05} noticed that the intracluster medium (ICM) shows elongated innermost X-ray isophotes, although they appear symmetric at larger radii. They observed a relatively high central entropy for the ICM, suggesting that A2631 is a galaxy cluster in the late stage of a merger. Similar conclusions were drawn by \cite{Zhang06} that classified A2631 as an ``offset centre'' cluster due to the non-concentric X-ray isophotes. They also confirmed the high mass of A2631, estimating $M_{500}=10.9\pm2.6\times 10^{14}$ M$_\odot$. All previous statements were also endorsed by high-resolution data observed by {\it Chandra} \citep{Mann12, Marrone12}.

Considering now the other end of the electromagnetic spectrum, observations taken by the Giant Metrewave Radio Telescope \citep[GMRT;][]{GMRT} at 610 MHz did not reveal any signal of extended radio emission in A2631 \citep{Venturi07}, which was subsequently ratified by \cite{Knowles19}. Radio halos are expected to appear as signatures of massive cluster mergers \citep{merging_book}. A2631 was also observed with the Sunyaev–Zel’dovich Array (SZA) by \cite{Reese12}, which estimated  $M_{500}\approx10^{15}$ M$_\odot$. Also using {\it Chandra} images, they corroborate the scenario of an elongated core both in X-ray and SZA data. However, the authors did not found any signals of substructures, which are proxies for disturbed systems \citep{Andrade-Santos12} nor a cool core, which is expected in relaxed clusters \citep[e.g.][]{Soja18}. The dynamical state of A2631 so far remains an unsolved puzzle.

Despite X-ray and radio exhibiting the most dramatic signatures of a cluster merger \citep{merging_book, markevitch_viki07},  this process leaves few  imprints in  the cluster optical properties \citep[e.g.][]{pinkney}. Within this context, \cite{Wen13} developed a new methodology to attest to a cluster's  dynamical state based only on the brightness distribution of member galaxies. They defined a relaxation parameter $\Gamma$ cut-off that satisfactorily separates the relaxed ($\Gamma \geq0$) and unrelaxed ($\Gamma<0$) systems with a success rate of 94 per cent. Unfortunately, they found an inconclusive classification for A2631 ($\Gamma=-0.02\pm0.10$).

The mass of a galaxy cluster is a fundamental parameter to probe theoretical models of large-scale structure formation and evolution as well as to constrain cosmological parameters \cite[e.g.][]{Kravtsov12, Pratt19}. Due to its unusual high mass, A2631 has been a popular target for mass surveys based on gravitational lensing. We can cite the  Local Cluster Substructure Survey \citep[LoCuSS\footnote{http://www.sr.bham.ac.uk/locuss/}; e.g.][]{Zhang08,Haines09} that has been analysing 50 of the most massive galaxy clusters in the local universe aiming to determine their masses as accurately as possible. Another remarkable example is the project Weighing the Giants \citep[WtG;][]{vonderLinden14, Applegate14}. The main advantage of lensing-based mass determinations is that they do not assume any prior about the cluster dynamical state, in contrast to  other techniques do. For example, X-ray hydrostatic mass estimators rely on the assumption of hydrostatic equilibrium in the innermost cluster region ($\sim0.2-1.25R_{500}$). Therefore the masses obtained will be highly biased in merging systems.

We present the multiwavelength mass estimates of A2631 in Table~\ref{tab:mass.comp}. A comparison of masses obtained from different techniques is an important tool to probe the dynamical state of a galaxy cluster \citep[e.g.][]{cypriano04}.  However, we observe a tension among the weak-lensing masses available in the literature. The values vary within a range of $\sim$[4--17] $\times10^{14} h_{70}^{-1}$ M$_{\odot}$  (disregarding the error bars) leading to a discrepancy of a factor of $4$. Such inconsistency can bias the statistic of cluster masses to cosmological purposes and affect the scaling relations with observables (e.g. SZ, X-ray, richness).

\begin{table}
\caption[]{Mass estimates of A2631 available in the literature. MRR stands for mass-richness relation,  SZ for Sunyaev-Zel'dovich and WGL for weak gravitational lensing. All values marked with $^\dagger$ were originally published in the form of $M_{500}$ and further translated into $M_{200}$ supposing that the halo masses follow an NFW density profile \citep{nfw96,nfw97} with concentration $c_{200}$ given by \cite{duffy08}.}
\setlength{\tabcolsep}{2pt}\centering
\begin{tabular}{lccc}
\hline \hline 
 Method & $M_{200}$                           & $R_{200}$  & Reference \\
        & $(10^{14} h_{70}^{-1}$ M$_{\odot})$ & (Mpc)      & \\
\hline 
caustic         &$3.77\pm0.66$            &  $1.07\pm0.08$ &  \cite{Geller13}\\[5pt] 
caustic         &$7.2\pm1.5^\dagger$              &  $1.70\pm0.12$    &  \cite{Maughan16}\\ 
\hline
dynamic       &$5.7\pm1.7$              &  $1.56\pm0.16$    &  \cite{Sifon16}\\
\hline
MRR             &$10.5$                   &  $1.9$         &  \cite{Bagchi17}\\ 
\hline
WGL             & $4.54_{-0.78}^{+0.89}$  & $1.45\pm0.10$  & \cite{Okabe10}\\[5pt] 
WGL             &$16.7\pm2.7$             &  $2.24\pm0.12$ &  \cite{Applegate14}\\[5pt] 
WGL             &$7.13_{-1.66}^{+2.07} $  &  $1.7\pm0.2$    &  \cite{Okabe16}\\[5pt] 
WGL             &$10.56^{+1.96}_{-2.04}$  &  $1.99_{-0.13}^{+0.11}$ &  \cite{Klein19}\\[5pt] 
\hline
SZ              & $9.2\pm1.9^\dagger$            &  $1.84\pm0.13$   &  \cite{Hasselfield13}\\ [5pt]
SZ              & $15.2\pm5.0^\dagger$            &  $2.2\pm0.3$   &  \cite{Planck13}\\ 
\hline
X--ray          & $16.9\pm3.9^\dagger$  & $2.4\pm0.2$  & \cite{Zhang06}\\[5pt] 
X--ray          & $9.7_{-1.8}^{+2.4\dagger}$     & $1.9\pm0.1$    &  \cite{Landry13}\\[5pt] 
X-ray           &$11.9\pm1.9^\dagger$             &  $2.00\pm0.11$    &  \cite{Maughan16}\\ 
\hline \hline 
\end{tabular}
\label{tab:mass.comp}
\end{table}

These inconsistencies encouraged us to conduct a comprehensive optical study of A2631 with the aim of (1) solving the discrepancy in the weak-lensing mass determination and (2) evaluating its current dynamical status. To reach these goals, we resorted to existing large field-of-view multiband images ($B$, $V$, and $R_C$) taken from the Subaru telescope archive as well as available redshift catalogues. With this wealth of data, we  mapped the spatial distribution of the cluster's dark matter and its galaxy content. We reconstructed the cluster mass field combining two different probes of the gravitational lensing effect, the shape distortion induced on the background galaxies and the change in their number counts. This combination contributed to a more precise determination of the cluster's total mass. We also quantified the  statistical significance among the spatial distribution of the cluster components, dark matter, galaxies, and gas. This piece of information works as a proxy for the dynamical state since a highly disturbed system can present a spatial detachment among these quantities \cite[e.g.][]{Massey11, merten, Pandge19, Monteiro-Oliveira20, Moura20}. Additionally, we determined the dynamical-based mass and searched for substructures as a probe for the cluster's dynamical state \cite[e.g.][]{Ribeiro13}.

The paper is organized as follows. In Section~\ref{sec:photo}, we present the cluster photometric analysis. The weak-lensing mass reconstruction is performed in Section~\ref{sec:wl}. Next, we describe A2631 from the dynamical point of view in Section~\ref{sec:dynamics}. The main findings of this work are discussed in Section~\ref{sec:discussion} and summarized in Section~\ref{sec:summary}.

Throughout this paper, we adopt the standard $\Lambda$CDM cosmology, represented by $\Omega_m=0.27$, $\Omega_\Lambda=0.73$, $\Omega_k = 0$ and $h = 0.7$. At the mean cluster redshift of $z = 0.2762$, we have a plate-scale of 1 arcsec equals 4.23 kpc \citep{CosmoCalc}.

\section{Photometric analysis}
\label{sec:photo}

\subsection{Imaging data}
\label{sec:imaging.data}

The galaxy cluster A2631 was observed in multifilters $B$, $V$, $R_C$ by the SuprimeCam\footnote{https://www.subarutelescope.org/Observing/Instruments/SCam/index.html} mounted at the Subaru telescope\footnote{https://www.nao.ac.jp/en/research/telescope/subaru.html} on 2004 July 18 ($V$ and $R_C$) and 2005 November 30 ($B$)\footnote{smoka.nao.ac.jp}. Details about the final products can be found in Table ~\ref{tab:imaging}.

\begin{table}
\begin{center}
\caption{Summary of the imaging data retrieved from the Subaru archive. Seeing was measured over a sample of bright and unsaturated stars within $19.5\leq{\rm mag}\leq 23.5$. The deepest $R_C$ band was chosen as the basis for the weak-lensing study.}
\begin{tabular}{lcc}
\hline
\hline
Band & Exposure time & Seeing  \\
     & (min)         & (arcsec) \\
\hline
$B$     & 12  & 1.0\\
$V$     & 18  & 0.8 \\
$R_C$   & 24  & 0.8 \\
\hline
\hline
\end{tabular}
\label{tab:imaging}
\end{center}
\end{table}

The image reduction was done with the standard semi-automated code {\sc SDFRED1} \citep{sdfred1}. The steps consisted of bias and overscan subtraction\footnote{These auxiliary images are also available at the SMOKA Science Archive.}, flat-fielding, atmospheric and dispersion correction, sky subtraction, auto-guide masking, and alignment (which was done simultaneously in all filters). Then, single images were combined and mosaicked into a final image for each filter. To ensure the exact correspondence among their Cartesian components ($x$, $y$), we performed their joint registration with {\sc IRAF}. The size of the final images was $37\times37.5$ arcmin$^2$, sufficient to map large projected distances from the centre of A2631 as required to perform a safe weak-lensing analysis.

Astrometric calibration was done based on the comparison of precise positions of bright and unsaturated stars from the ``Fourth US Naval Observatory CCD Astrograph Catalog'' \citep[UCAC4;][]{Zacharias13} and in our images. After this process, we found a positional rms of $0.19\pm 0.18$ arcsec, thus ensuring a satisfactory accuracy for our forthcoming analysis.

In order to perform the photometric calibration, i.e., to transform the measured fluxes into the AB magnitude system \citep{Oke74}, we need to observe some standard field (stars with known AB magnitudes) over several (or at least two) air masses on the same date as the science observations. For our data set, only the filter $R_C$ matches this requirement. We determined the extinction coefficients $k^\prime$ and $k^{\prime\prime}$ through the following relation,
\begin{equation}
    m_{\rm ref}+m_{\rm inst} = m_0^{\rm cal}+k^\prime X+k^{\prime\prime}C
    \label{eq:calib.ext}
\end{equation}
with $m_{\rm ref}$  being the reference AB magnitude, $m_{\rm inst}$ the instrumental magnitude, $X$ the airmass, $C$ the AB index colour (e.g. $V-R_C$), and $m_0^{\rm cal}$ a calibration constant. Then a low-exposure science image was calibrated also taking into account the factor $2.5\log\tau$ where $\tau$ is the ratio between the time exposure of the science (240 s) and the standard star (10 s) images. To estimate $C$, we adopted the index colour of the galaxy type $Sab$  located at $z\sim 0.8$ \citep{fukugita}. Then we compared this calibrated image with the co-added $R_C$ image. We found the magnitude zero-point  $R_C^0=33.68\pm0.03$.

To calibrate the filter  $V$, we turn to SDSS DR12 \citep{Alam15} to find the calibration coefficients $\xi$ and $\epsilon$  for the single standard field observed, as given by the linear fit
\begin{equation}
   V=g^\prime+\xi(g^\prime-r^\prime)+\epsilon.
       \label{eq:calib.sdss}
\end{equation}
Following this, $\xi$ and $\epsilon$ were applied to calibrate a low-exposure science image that served as the base to the photometric calibration of the co-added $V$ image. The magnitude zero-point obtained is $V^0=32.611\pm0.003$. A similar procedure was done for $B$ filter, except that we did not have any standard field from the same date as the science observations. We overcome this issue obtaining  $\xi$ and $\epsilon$ by comparing the SDSS DR12 catalogue with the  $B$ catalogue of A2034 \citep{Monteiro-Oliveira18}. We found  $B^0=32.658\pm0.006$.

Photometric catalogues were created using {\sc SExctractor} \citep{sextractor} in double mode with the deepest $R_C$ image as a base. Galaxies were identified according to two complementary criteria. For the brightest objects ($R_C<19.5$) galaxies  will correspond to ${\rm CLASS\_STAR}<0.8$ whereas for $R_C \geq 19.5$ their full width at half-maximum (FWHM) should be greater than  $0.9$ arcsec. This value is $0.1$ arcsec larger than the seeing to ensure the selection of well resolved objects.

\subsection{Identification of the red-sequence}
\label{sec:red.seq}

The central region of rich galaxy clusters is inhabited preferentially by red members \citep{dressler80}. As a result of the homogeneous photometric properties, these galaxies occupy a well defined {\it locus} in a colour-colour map \citep[CC; e.g.][]{Medezinski18}.  Here, we identified the red-sequence {\it locus} of A2631 on the $B-V$ versus $V-R_{C}$ space applying the statistical subtraction method \citep[see][for more details]{Monteiro-Oliveira17b}.

\begin{figure}
\begin{center}
\includegraphics[width=\columnwidth, angle=0]{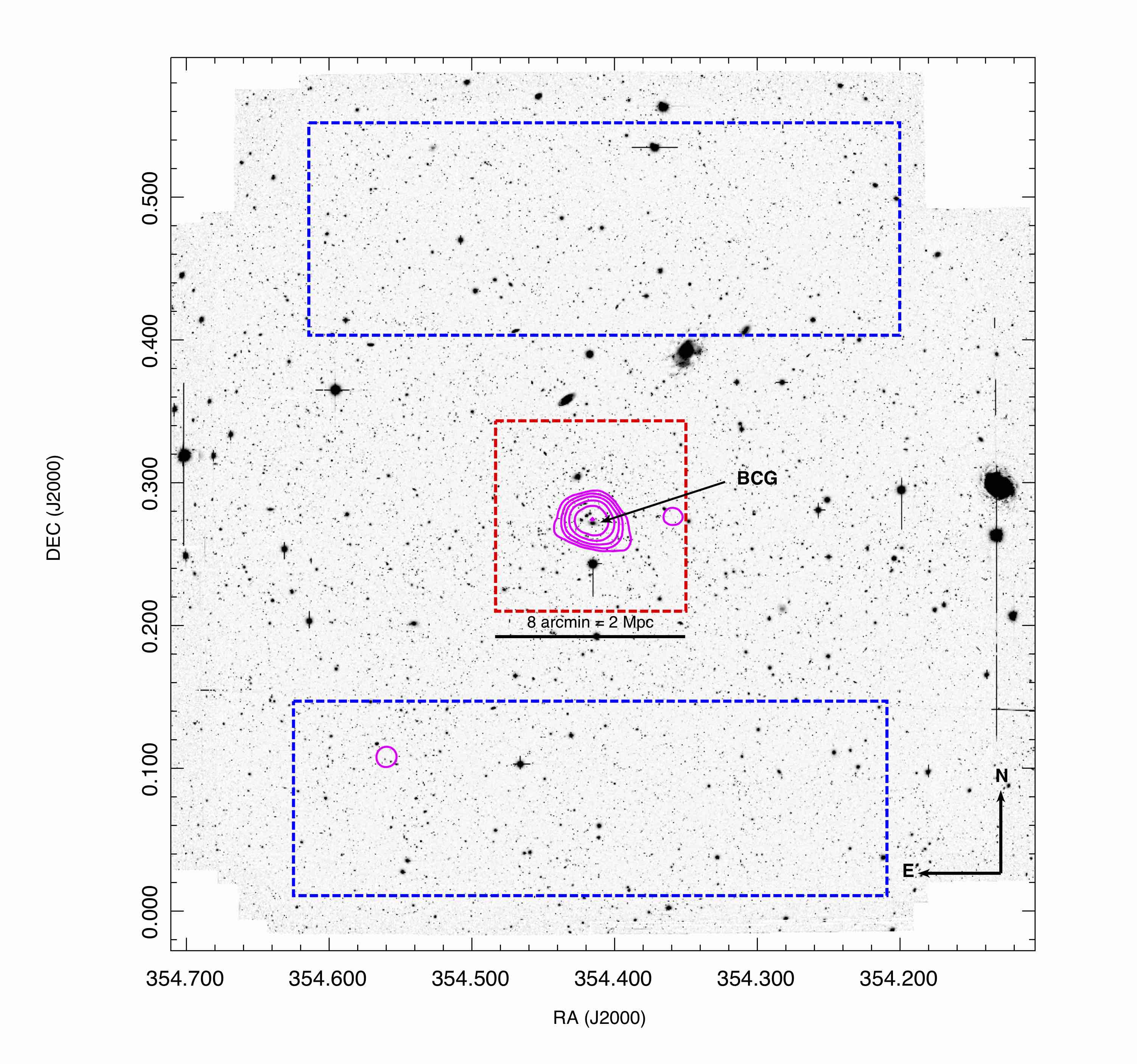}
\caption{Subaru Suprime-Cam $R_C$ image of the A2631 field. Overlaid magenta contours show the luminosity map (i. e. the projected density weighted by the $R_C$ flux) of the identified cluster red-sequence galaxies. The boxes define the regions considered for the statistical subtraction process applied to identify the red-sequence galaxies. Blue boxes enclose regions supposed to be dominated by field galaxies in contrast with the inner region (red box)  where the cluster member counts will prevail.}
\label{fig:field}
\end{center}
\end{figure}

We selected 964 photometric members over the field within $R_C<23$. This corresponds to the faintest limit where galaxy counts in the innermost region are higher than those in the outskirts, where field counts are expected to dominate. The red-sequence projected density weighted by the $R_C$ flux  is shown in Fig.~\ref{fig:field}. The map shows the results of the smoothing of photometric members inside  a $5$ arcsec$^2$ cell by an Epanechnikov kernel with a scale of 70 arcsec.

The luminosity spatial distribution is dominated by the brightest cluster galaxy (BCG) and some luminous neighbours. A2631 is well described by a single structure just as the majority of relaxed clusters are \citep{Wen13}. Next, we will recover the cluster projected mass distribution in order to check if this scenario is supported.

\section{weak-lensing analysis}
\label{sec:wl}

\subsection{Basic concepts}
\label{sec:comncepts}

We can approach the effects of the gravitational lens from the projection of its scalar potential,
\begin{equation}
\psi (\bm{\theta})= \dfrac{1}{\pi}  \  \int d^2\theta' \ \kappa(\bm{\theta'}) \ln|\bm{\theta} - \bm{\theta'}| \ \mbox{.}
\label{eq:potential}
\end{equation}
This is conveniently defined so that we can directly write
\begin{equation}
\nabla^2 \psi = 2\kappa \ \mbox{,}
\label{eq:kappa.potential} 
\end{equation}
with
\begin{equation}
\kappa(\bm{\theta}) \equiv \dfrac{\Sigma(\bm{\theta})}{\Sigma_{\rm cr}}
\label{eq:kappa}
\end{equation}
being the projected mass density of the lens also known as convergence. It is written in units of the lensing critical density 
\begin{equation}
\Sigma_{\rm cr} = \dfrac{c^2}{4\pi G}\dfrac{D_s}{D_{ds} D_d}
\label{eq:sigmacr}   
\end{equation}
where $D_s$, $D_{ds}$, and $D_d$ are, respectively, the angular diameter distances to the source, between the lens and the source, and to the lens.

Besides the scalar convergence, the gravitational lensing field can be described by a  spin-2 tensor, the shear
\begin{equation}
\gamma = \gamma_1+i\gamma_2 \ \mbox{,}
\label{eq:shear}   
\end{equation}
whose components are both second derivatives of the projected gravitational potential (Equation~\ref{eq:potential}). An alternative way to define these quantities is in terms of their tangential component to the lens centre $\gamma_+$ and another one $45^\circ$ in relation to that, $\gamma_\times$.

In the absence of any gravitational lens, the  value of the averaged ellipticities $\langle e \rangle $ of background\footnote{With respect to the targeted galaxy cluster.} galaxies is expected to be zero. However, the lens effect acts to induce a coherent distortion whose averaged ellipticity will tend to the effective shear $g$:
\begin{equation}
\langle e \rangle \simeq g \equiv \frac{\gamma}{1-\kappa} \ \mbox{.}
\label{eq:red.shear}   
\end{equation}
As $\gamma$, the full ellipticity, and the effective shear are both a spin-2 tensor. Mathematically the weak regime corresponds to $\kappa \ll 1$. In this case, we have $g \approx \gamma$.

During the passage through the gravitational lens, there is  conservation of the angular momentum and energy of the photons coming from the background galaxies. Considering also the absence of emitters and absorbers in the path of the light beam from the source to the observer, we conclude that there is numeric  conservation of the photons. From Liouville's theorem, we can enunciate an important characteristic of the phenomenon of gravitational lensing: the conservation of  surface brightness. Due to the amplification of the image size, the observed flux (superficial brightness $\times$ the image's solid angle) will be increased by the same factor implying that the lensed image will be brighter than its source. This increase is quantified by the magnification $\mu$:
\begin{equation}
\mu = \frac{1}{(1-\kappa)^2 - |\gamma|^2}\ \mbox{.}
\label{eq:mu}   
\end{equation}

Parallel to the reconstruction of the mass distribution of the measured lens from the measurement of the distortion caused in the background galaxies, it is possible to do the same work from the measurement of the magnification effect caused locally in the spatial distribution of background galaxies.

The so-called ``magnification bias'' is the joint manifestation of two effects of the magnification phenomenon: at the same time that it increases the flux from the source, thus allowing the detection of intrinsically weaker objects, it also magnifies by the same value the element of the projected area of the sky acting to decrease the apparent density of objects.

The magnification bias value is related to both the magnification factor $\mu$ (Equation~\ref{eq:mu}) and the slope
\begin{equation}
\alpha = \dfrac{d \log {\rm N (<m)}}{dm}
\label{eq:slope}
\end{equation}
of the intrinsic relationship between the logarithm of the galaxy counts as a function of their magnitudes, measured in a field unaffected by the effect of the lens.

When approaching for a circular lens, the numerical radial density of background galaxies can be written as
\begin{equation}
N(<m,r)=N_0(<m)\mu(r)^{2.5\alpha-1}\ \mbox{,}
\label{eq:bias.mag}
\end{equation}
where $ N_0 (<m) $ is the intrinsic density of objects measured in a region far enough away to not be affected by the gravitational lens.

The magnification bias $N(<m, r)$ is present only if $\alpha \neq0.4$, in which case the numerical increase of the magnified sources is exactly compensated by the apparent expansion of space. For $\alpha>0.4$ we will see an increase in the density of galaxies while in the $\alpha< 0.4$ regime there will be a decrease in this amount compared to $N_0(<m)$.

Since the magnification bias is not based on any measurement of shape in the galaxies nor does it require knowledge about the original shape, its use instead of the technique based on the measurement of distortion would be obvious. However, in the weak-lens regime ($ \kappa \approx |\gamma|$), the ratio between the signal and the noise of both extensions, considering $\sigma_e = 0.3$ and $\alpha=0.2$,
\begin{equation}
R_{\rm s/m}=\dfrac{|\gamma|}{\sigma_e}\dfrac{1}{\kappa(5\alpha-2)}\approx3\mbox{,}
\label{eq:sn}
\end{equation}
favours analysis based on distortion in the shape of galaxies \citep{mellier99}.

Despite this limitation, mass reconstruction through magnification bias constitutes a test to check consistency in the measurement of masses as this approach to the phenomenon of gravitational lenses is not susceptible to systematic effects from the measurement of the shape of galaxies and point spread function (PSF) correction.

\subsection{Shear data set}
\label{sec:shear}

\subsubsection{Source selection}
\label{sec:source.shape}

We refer to the background, i.e. those galaxies located at higher redshift than A2631, as the sources because they are the basis for shape and/or  numerical density measurements required by the weak gravitational lensing technique. Despite the source galaxies being spatially spread along the field, we can resort to the CC map to identify the {\it loci} where the contamination by both foreground and red-sequence galaxies is lower \citep[e.g.][]{capak07,med10}.

\begin{figure}
\begin{center}
\includegraphics[width=\columnwidth, angle=0]{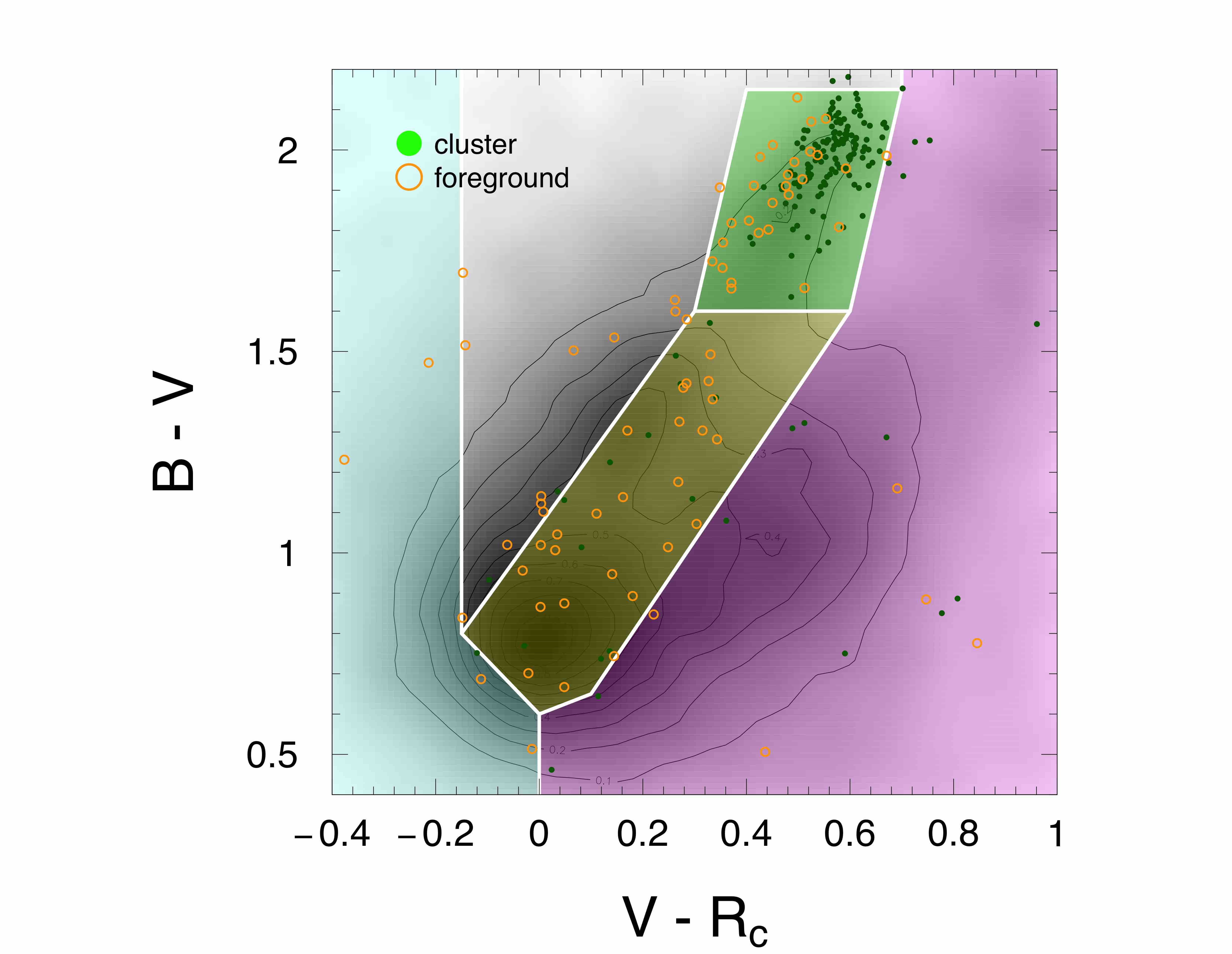} 
\caption{Galaxy populations in the colour-colour diagram. The  {\it loci} preferentially occupied by red-sequence (found after background subtraction) and  foreground galaxies correspond, respectively, to the green and yellow polygonal regions. The orange circles and green dots are, respectively, cluster and foreground galaxies, according to their redshifts. The {\it locus} dominated by the source galaxies is subdivided in two: the red background (magenta region) and the blue background (cyan region).} 
\label{fig:locus}
\end{center}
\end{figure}

Before identifying the source {\it locus} in the  $B-V$ versus $V-R_C$ space, we need to draw the foreground location. According to synthetic models of galaxy evolution \citep[e.g.][]{Medezinski18}, they tend to be bluer than the red-sequence and form a dense cloud in the CC map. We show, in Fig.~\ref{fig:locus}, the position of the foreground {\it locus} after a visual inspection. As a sanity check, we overlaid our spectroscopic sample, which we will describe in more detail in Sec.~\ref{sec:dynamics}.

We then selected the background {\it locus} excluding both red-sequence and foreground regions. Background galaxies are usually classified into blue ($V-R_C\leq0$) or red populations \citep[$V-R_C>0$; e.g.][]{Medezinski18}. This taxonomy is relevant when we are interested in the magnification bias effect, i.e the change in the source counts caused by the gravitational lens \citep[e.g.][]{Monteiro-Oliveira17a}. For a shear-based analysis, both samples are indistinguishable.

\subsubsection{Shape measurements}
\label{sec:source.shape.mea}

The PSF is the combined effect of the atmospheric blurring plus the response of the telescope optics and instrumentation on the image. Fortunately, these effects can be described mathematically leading to an analytical expression of the PSF effects. To build this expression, we identified and analysed the shape of bright and unsaturated stars across the $R_C$ image.  They should be exact point-like sources in the absence of the PSF effects and they are, without doubt, unlensed objects.

\begin{figure}
 \begin{center}
\includegraphics[width=\columnwidth, angle=0]{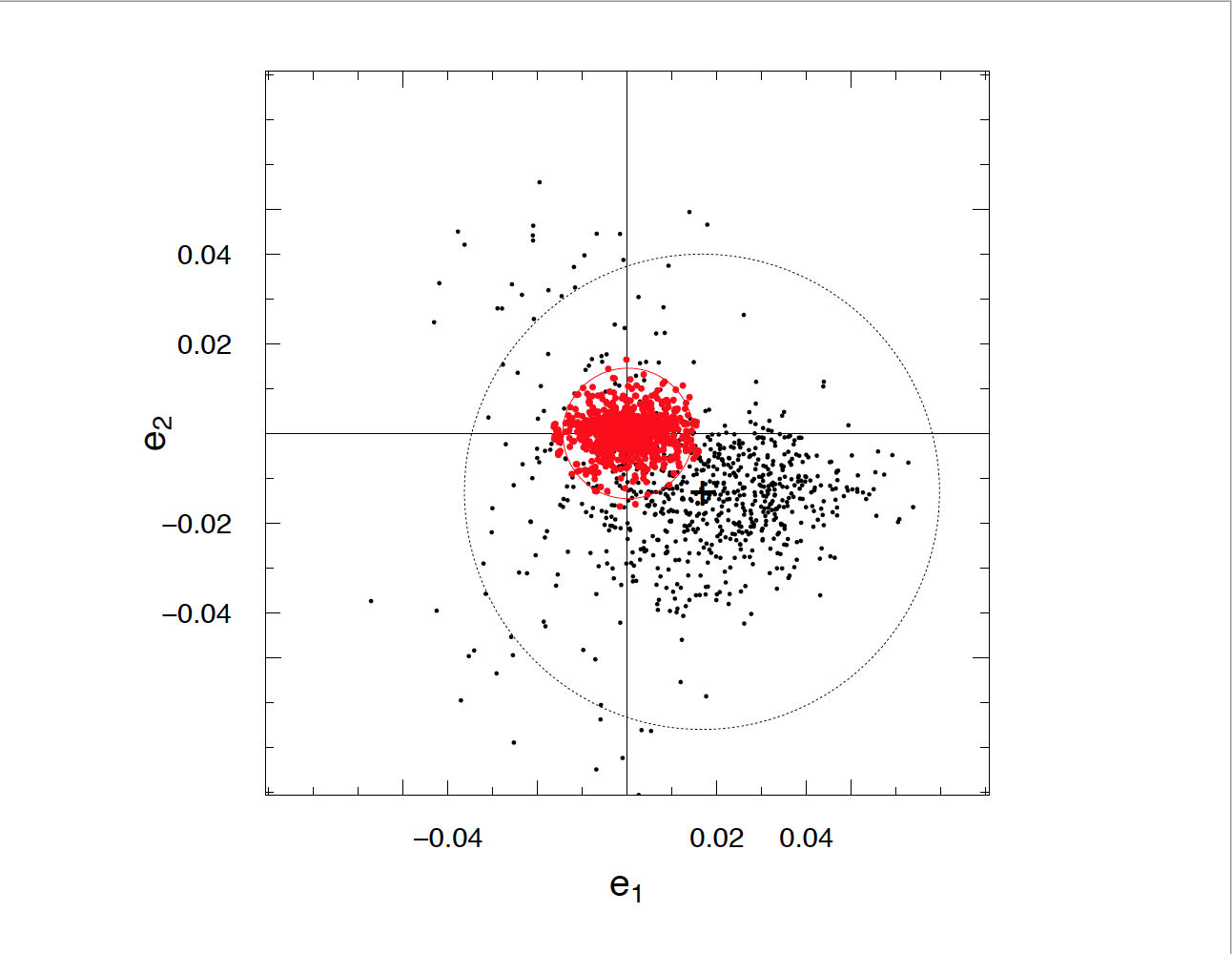} 
\caption{PSF ellipticity. Black dots show the raw components $e_1$ and $e_2$ measured from bright and unsaturated stars. Red dots correspond to the residual between the original ellipticities and the analytical PSF function. Dashed circles enclose 95 per cent of the data. The averaged residuals are $0.0002\pm 0.0069$ for $e_1$ and $ -0.00002 \pm 0.00444$ for $e_2$.} 
\label{fig:psf.stars}
\end{center}
\end{figure}

In order to model the shape parameters (ellipticity components $e_1$ and $e_2$ plus the FWHM), we resort to the Bayesian code {\sc im2shape} \citep{im2shape}. It models star profiles as single Gaussians and, in this case, it does not do any  PSF deconvolution. Following this,  the discrete values of parameters were spatially interpolated to create a continuous function across the image. This is done by {\sc thin plate regression} \citep[tps;][]{fields} in the {\sc R} environment. Aiming to remove obvious outliers, we reapplied {\sc tps}  three times;  in each iteration, we removed the objects with the 10 per cent largest absolute residuals. We present, in Fig.~\ref{fig:psf.stars}, the measured stellar ellipticities and their respective residuals.

The averaged ellipticity is $\langle e_1 \rangle=0.014\pm0.021$ and  $\langle e_2 \rangle=-0.012\pm0.018$. As in practice, stars are not point sources, all deviations from null values are due to the PSF effect. As illustrated by Fig.~\ref{fig:psf.stars}, our model showed a good match with the data as attested by the very low residuals, $0.0002\pm 0.0069$ for $e_1$ and $ -0.00002 \pm 0.00444$ for $e_2$.

Now, we should measure the shape parameters of the source galaxies and perform PSF deconvolution to extract the weak-lensing signal. There is also the effect of the unknown original shapes, but this will be treated later. We  again use the code {\sc im2shape} but now it models the galaxies as a sum of Gaussians with an elliptical basis and performs local PSF deconvolution. The final result is a PSF-free catalogue of the galaxy ellipticities $e_1$ and $e_2$ and the respective uncertainties $\sigma_e$. In addition, we removed all source galaxies with $\sigma_e>2$ or those showing evidence of blending. Our final source catalogue is composed of 15153 galaxies, leading to a projected density of 13.6 galaxies per arcmin$^2$.

In the weak-lensing context, observational parameters are translated into physical quantities through the critical surface density $\Sigma_{\rm cr}$. As pointed out by Equation~\ref{eq:sigmacr}, we need to know the distribution of the source's redshift. So, we applied the same colour and magnitude cuts (Sec.~\ref{sec:source.shape}) in the photometric redshift catalogue of the Cosmic Evolution Survey \citep[COSMOS;][]{Ilbert09}. We found $\Sigma_{\rm cr}=2.80\pm0.09\times10^9$ M$_\odot$ kpc$^{-2}$ and a mean redshift of $z_{\rm source}\simeq1.1$.

\subsubsection{The projected mass distribution}
\label{sec:proj.mass}

Source galaxies have an intrinsic unknown ellipticity that makes them noisy tracers of the shear field. To take this fact into account, we averaged the quadratic sum of the ellipticity components $e_1$ and $e_2$ in the image outskirts where we suppose the lensing signal is insignificant. We found $\sigma_{\rm int}=0.45$ and therefore consider it as the intrinsic error on the ellipticities.

The field projected mass distribution was recovered by the maximum entropy algorithm \citep{seitz98} implemented in the Bayesian code {\sc LensEnt2} \citep{LensEnt2}. The code finds the best solution through the maximization of the evidence based on the comparison between the shear field sampled by source ellipticities with those predicted by the model. Since each galaxy ellipticity induced by the gravitational lens is correlated with the neighbourhood, we should adopt a smoothing scale. This also takes into account the fact that each galaxy is a noisy probe of the shear field. The smoothing is implemented in the code by the intrinsic correlation function (ICF), chosen as a Gaussian filter by us.

\begin{figure}
 \begin{center}
\includegraphics[width=\columnwidth, angle=0]{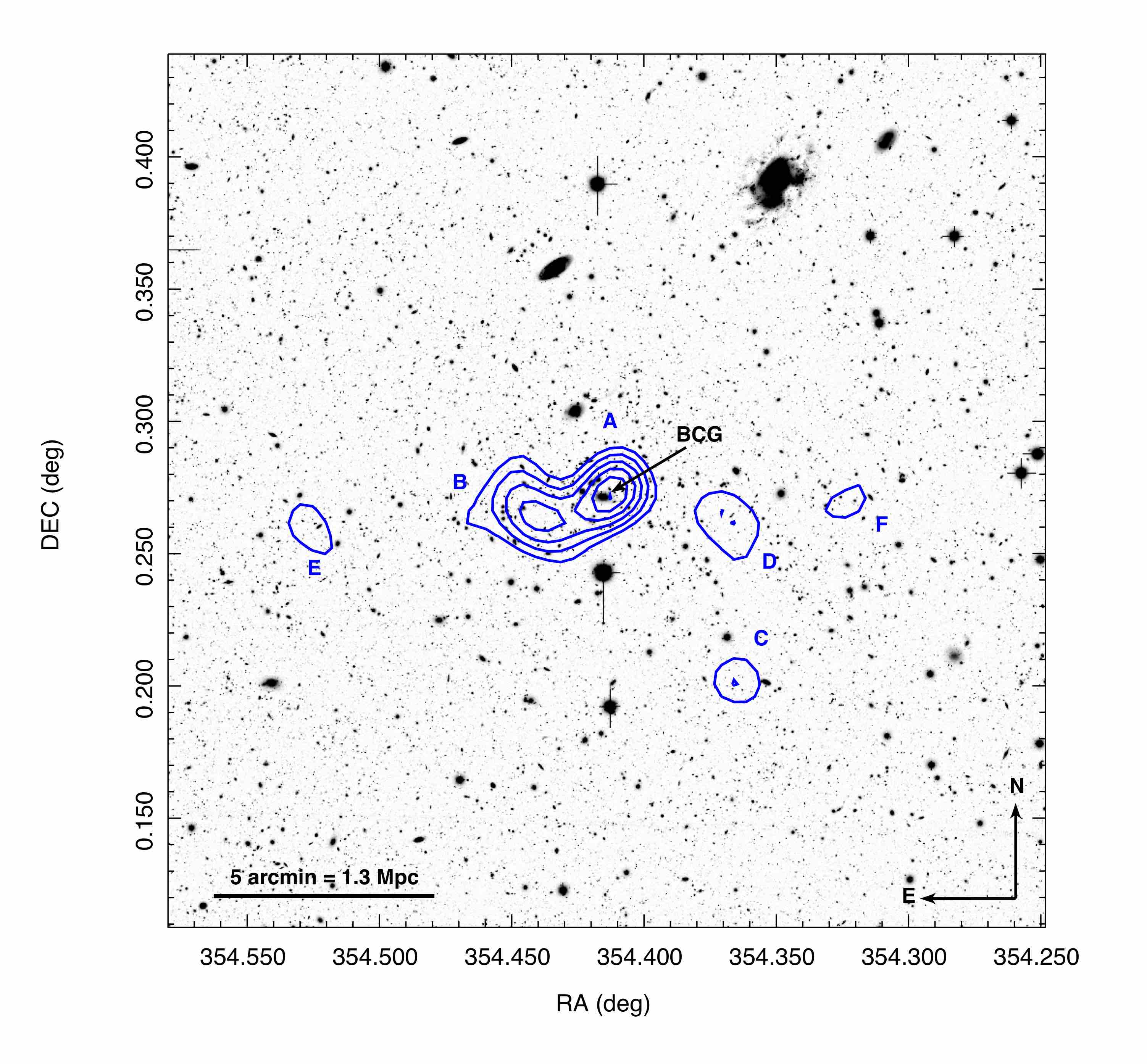}
\caption{Convergence map in the A2631 region overlaid on the Subaru $R_C$ image. The blue convergence contours, which are directly related to the projected mass distribution, start from $6\sigma_\kappa$ with $\sigma_\kappa$ being the noise level, and are in units of $\sigma_\kappa$. The galaxy cluster A2631 appears as a mass concentration surrounding the BCG (halo A). A companion structure seems to appear on the eastern side (halo B); this will be a matter for further discussion in our mass modelling as well as the other four mass peaks found (C--F).} 
\label{fig:mass.map}
\end{center}
\end{figure}

To find the optimized $\sigma_{\rm ICF}$, we made mass reconstructions within the interval [60,120]~arcsec. For each, we searched for peaks in mass to compute the statistic of the numeric detections in the function of their significance. We found that peak detections above $6\sigma_\kappa$ remain almost constant for $\sigma_{\rm ICF}\geq 80$ arcsec; therefore this value was adopted. The noise level in the convergence map, $\sigma_\kappa$, was obtained after performing 100 realizations of the shear field without the cluster lens signal. For that, each galaxy ellipticity was rotated  by a random angle in the interval [0,180[.  We present our fiducial convergence map in Fig.~\ref{fig:mass.map}. The main characteristics of the convergence map are shown in Table~\ref{table:mass.charac}.

\begin{table}
\begin{center}
\caption[]{Relevant quantities of the weak-lensing convergence map.}
\begin{tabular}{l c}
\hline
\hline
$N_{\rm g}$ (gal. arcmin$^{-2}$)  &  13.6\\ 
ICF FWHM  (arcsec) &  80 \\
$\sigma_\kappa {}^\star$  & 0.035\\
\hline
\hline
\end{tabular}
{\small  \item  ${}^\star$ Noise level in the convergence map.}
\end{center}
\label{table:mass.charac}
\end{table}

The most noticeable feature in the convergence map is the presence of a significant halo matching the BCG position (halo A in Fig.~\ref{fig:mass.map}). This structure is nearly aligned with the red-sequence spatial distribution (Fig.~\ref{fig:field}) which allows us to recognize it as the halo of A2631. Also remarkable is the presence of a structure located $\sim$1.5 arcmin east of the BCG (halo B). Checking Fig.~\ref{fig:field}, we did not see any relevant members related to it suggesting that this structure does not belong to our target cluster. Anyway, we will consider this ``companion'' structure in our forthcoming mass modelling.

\subsection{Magnification bias data set}
\label{sec:mag.bias}

\subsubsection{Source selection and counts}
\label{sec:mag.source}

Following \cite{Monteiro-Oliveira17a}, we considered only red sources to describe the lens-induced magnification bias. This subsample has a logarithmic slope (Equation~\ref{eq:slope}) computed at the image outskirts \footnote{At least 10 arcmin away from the cluster centre traced by the BCG.} $\alpha= 0.165\pm0.013$ near the completeness limit ($R_C = 25.4$). Thus, we expect the lensing effect to cause a depletion of the number counts. For this sample, the critical surface density for this sample was estimated in the same way as for the shear data set (Sec.~\ref{sec:source.shape}) leading to $\Sigma_{\rm cr}=2.91\pm 0.13\times10^9$ M$_\odot$ kpc$^{-2}$ and a mean redshift of $z_{\rm source}\simeq 0.86$.

To map the magnification bias effect, we used the ``count-in-cells'' technique. We divided our image into $29\times35 = 1015$ squared cells with 1 arcsec per side. Additionally, we masked out regions occupied by large objects as saturated stars and galaxies (including the BCG). Then, we computed the area of each cell after discounting the masked regions. Finally, the counts in regions away from the BCG ($> 10$ arcmin) yielded a baseline number count of $N_0=10.7\pm 4.2$ galaxies per arcmin$^{-2}$.

\subsection{Mass modelling}
\label{sec:mass.modeling}

The lensing observables (shear and galaxy counts) were modelled as if they were induced by mass halos following an NFW profile \citep{nfw96,nfw97}. For a given set of NFW profile parameters, the shear field was built following \cite{wright00} prescriptions. The theoretical profile is defined by four parameters: the lens centre coordinates $x,y$, the mass enclosed within a radius where the density is 200$\times$ the critical density of the Universe\footnote{$\rho_c=\frac{3H^2(z)}{8\pi G}$,} $M_{200}$, and the dimensionless halo concentration $c_{200}$. To focus on the determination of the halo centre and its mass, we opt to fix the concentration by adopting the $M_{200}-c_{200}$ relation presented by \citet{duffy08},
\begin{equation}
c_{200}=5.71\left(\frac{M_{200}}{2\times10^{12}h^{-1}M_{\odot}}\right)^{-0.084}(1+z)^{-0.47}.
\label{eq:duffy_rel}
\end{equation}
where $z$ is the cluster redshift.

We are looking for the best model to describe the observed mass distribution in the A2631 field. For this, we computed three different models, described in Table~\ref{tab:model.desc}. These models were designed to confirm or not whether the other mass clumps found (see in Fig.~\ref{fig:mass.map}) correspond to substructures of A2631.

\begin{table*}
\begin{center}
\caption{Description of the models proposed to describe the mass distribution in the A2631 field. $\theta$ refers to the vector parameter of each model, with $M=M_{200}$ and $x,y$ the coordinates of the respective halo centre.}
\begin{tabular}{lcc}
\hline
\hline
Model & Haloes & $\theta$\\
\hline  
\#1 & 1 (A)    & $M_A$, $x_A$, $y_A$ \\
\#2 & 2 (A, B) & $M_A$, $M_B$, $x_A$, $y_A$, $x_B$, $y_B$ \\
\#3 & 6 (A--F) & $M_A$, $M_B$, $M_C$, $M_D$, $M_E$, $M_F$, $x_A$, $y_A$, $x_B$, $y_B$ \\
\hline
\end{tabular}
\label{tab:model.desc}
\end{center}
\end{table*}

We should consider that each source galaxy is simultaneously affected by the $N$ computed halos. In this case, we write the effective quantities for a considered galaxy as
\begin{equation}
\kappa = \sum_{i=1}^{N}\kappa_i \ ; \ \gamma_j = \sum_{i=1}^{N}\gamma_{j,i},
\label{eq:kappa_eff}
\end{equation}
with  $j={1,2}$.

For the shear data set we can write the $\chi^2$ statistic:
\begin{equation}
\chi^2_s=\sum_{j=1}^{N_{{\rm sources}}} \sum_{i=1}^{2}  \frac{[g_i(M_{200},x,y)-e_{i,j}]^ 2}{\sigma_{\rm int}^2+\sigma_{{\rm obs}_{i,j}}^2}, 
\label{eq:chi2.shear}
\end{equation}
where $g_i(M_{200}, x, y)$ is the predicted reduced shear (Equation~\ref{eq:red.shear}), $e_{i,j}$ is the measured ellipticity and $\sigma_{{\rm obs}_{i,j}}$ is the shape error given by {\sc im2shape}.

The log-likelihood for the shear data set can be written as
\begin{equation}
\ln \mathcal{L}_s \propto - \frac{\chi^2_s}{2}. 
\label{eq:log.lik.shear}
\end{equation}

From the magnification bias view, we can measure the lensing signal by comparing the measured counts with the theoretical prediction as 
\begin{equation}
 \chi^2_m=\sum_{i=1}^{N_{\rm cells}} \frac{[N_i-N_0~\mu(M_{200},x,y)^{2.5\alpha-1}]^2}{\sigma_{N_0}^2}  \frac{W^2_i}{\sum_{j=1}^N{W^2_j}},
\label{eq:chi2.mag}
\end{equation}
where $N_i$ corresponds to the cumulative counts in each cell corrected by the unmasked cell area and $W = \sqrt{1-A_{\rm mask}/A_{\rm total}}$ is a weight that penalizes cells with small effective areas. Then, the log-likelihood is 
\begin{equation}
\ln \mathcal{L}_m \propto - \frac{\chi^2_m}{2}. 
\label{eq:log.lik.mag}
\end{equation}
%

Following a Bayesian approach, we considered two additional   ``nuisance parameters'' in the models, $N_0$ and $\alpha$, related to the counts of the unlensed population (Sec.~\ref{sec:mag.source}). They will be fitted along with the halo-related parameters, but we established normal priors for both based on our measurements. For the masses, we applied a flat prior $0<M_{200}\leq 1\times10^{16}$ M$_\odot$, which avoids the consideration of unrealistic values and thus accelerate the convergence of the model. The same strategy was adopted for the halo centres with a prior $(x - x_c)^2 + (y - y_c)^2 \leq 80$ arcsec ($\sim$344 kpc), where $x_c, y_c$ are the halo centre coordinates.

After these considerations, we can write the posterior of our problem as 
\begin{multline}
\noindent \mathcal{P}(\theta,N_0,\alpha|{\rm data}) \propto  \\
\mathcal{L}_s({\rm data}|\theta) \times \mathcal{L}_m({\rm data}|\theta,N_0,\alpha)~\Pi(N_0)~\Pi(\alpha)\, .
\label{eq:posterior}
\end{multline}
where $\theta$ is the vector of parameters.

\subsection{Results}
\label{sec:results}

The source galaxies considered in our models were restricted to those contained in an area of $15\times 15$ arcmin$^2$ centred on the BCG. Then, the posterior described in Equation~\ref{eq:posterior} was sampled for each model by the MCMC algorithm with a Metropolis sampler {\sc MCMCmetrop1R} \citep{MCMCpack}. We generated four chains with $10^{15}$ elements allowing an additional chain of $10^4$ first points in each as ``burn-in''. 

For the best model selection, we computed the Bayesian information criterion (BIC),
\begin{equation}
{\rm BIC}= k\ln n-2\ln \mathcal{L}\,
\label{eq:BIC}
\end{equation}
with $k$ being the number of model parameters, $n$ the number of data points, and $\mathcal{L}=\mathcal{L}_s+\mathcal{L}_m$ is the maximized value of the model's likelihood.  Among a finite number of models, those with the lowest BIC will be preferred. In Table~\ref{tab:BIC}, we present a statistical comparison of the models considered.

\begin{table}
\begin{center}
\caption{Comparison of the models based on the BIC statistics. For the sake of comparison, we show the results for the individual data sets: shear (s), magnification (m), and both combined (s+m) .}
\begin{tabular}{lcccc}
\hline
\hline
Model & Parameters& \multicolumn{3}{c}{$\Delta$BIC with respect to model \#1}\\
&  & s & m & s+m \\
\hline  
\#1 & 3  & 0 & 0 & 0 \\
\#2 & 6  & 22 & 21 &20 \\
\#3 & 10 & 48 & 55 &43 \\
\hline
\end{tabular}
\label{tab:BIC}
\end{center}
\end{table}

Concerning the BIC criterion, the lowest index is presented by the simplest model that describes a single-halo at A2631's location. This description is strongly preferred in relation to the others as indicated by the large value of $\Delta$BIC  \citep{kass95}. Consequently, this will be our fiducial model hereafter.
   
\begin{figure*}
 \begin{center}
\includegraphics[width=\textwidth, angle=0]{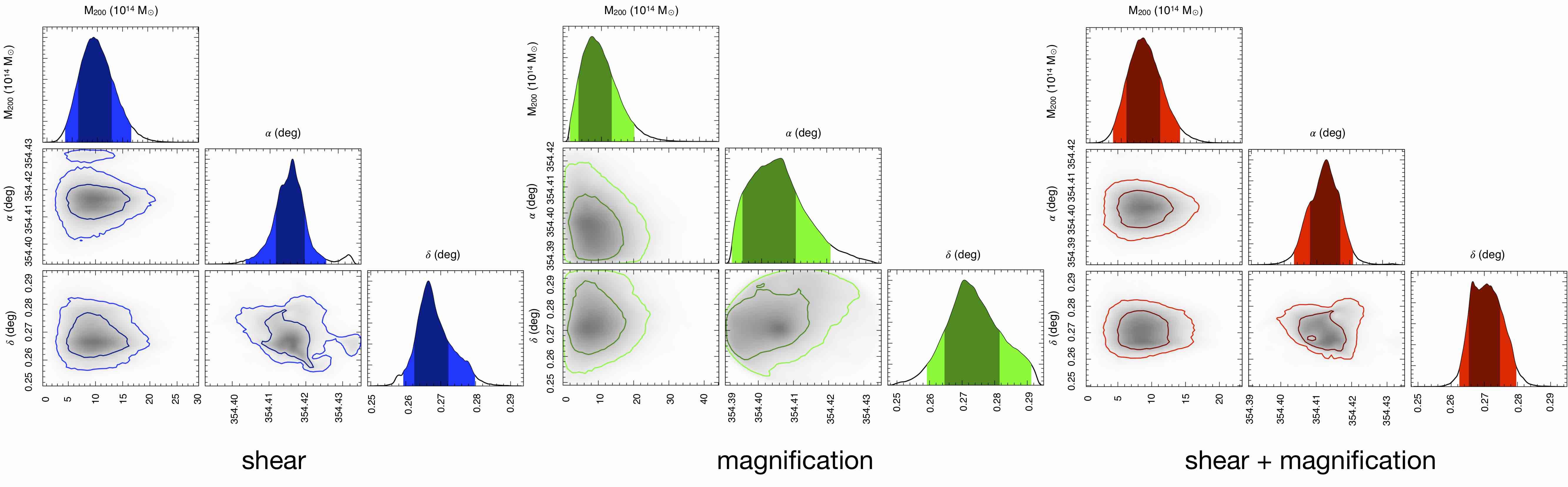}
\caption{Posterior mapped by the MCMC-based analysis for the data sets considered in our single-halo model. We translated the Cartesian coordinates $x$, $y$ to $\alpha$, $\delta$. In the diagonals, we present the marginalized posterior of the respective parameters.} 
\label{fig:model.dm.post}
\end{center}
\end{figure*}

The parameter estimation based on the analysis of the combined shear and magnification data sets is presented in Table~\ref{tab:masses.dm}. For the sake of comparison, we also show the estimation based on the individual data sets. We considered at face value the median of the respective posterior marginalized over all other parameters (Fig.~\ref{fig:model.dm.post}). The error bars correspond to the 68 per cent range of the MCMC samples.

\begin{table}
\begin{center}
\caption{Parameter estimation according to the single-halo model. ``s'' stands for the analysis considering only the shear data set, ``m'' for that considering only the magnification bias data set and ``s+m'' shows the results for the combined data sets.}
\begin{tabular}{lcccc}
\hline
\hline
 & $M_{200}$                      & $R_{200}$       & $\alpha$       & $\delta$     \\
 &  ($10^{14}$ M$_\odot$)         &           (Mpc) &          (deg) &          (deg)\\
 
\hline  
s       & $9.8_{-3.7}^{+3.0}$  &  $1.9_{-0.3}^{+0.2}$ & $354.416_{-0.004}^{+0.004}$ & $0.268_{-0.005}^{+0.005}$\\ [5pt]
m       & $9.0_{-6.0}^{+4.2}$  &  $1.8_{-0.6}^{+0.3}$ &  $354.404_{-0.010}^{+0.006}$ & $0.273_{-0.008}^{+0.009}$\\ [5pt] 
s+m     & $8.7_{-2.9}^{+2.5}$  &  $1.8\pm0.2$ & $354.412_{-0.005}^{+0.005}$ & $0.271_{-0.006}^{+0.004}$\\
\hline
\hline
\end{tabular}
\label{tab:masses.dm}
\end{center}
\end{table}

According to our modelling, the galaxy cluster A2631 has a mass of $M_{200}=8.7_{-2,9}^{+2.5}\times10^{14}$ M$_\odot$. From Table~\ref{tab:masses.dm}, we confirm that the analyses based on independent shear and magnification data sets are consistent with each other, having a significant overlap within 1$\sigma$. The combination of both decreased the error bars by $\sim$20 per cent in comparison to the shear data set only.

\begin{figure}
\begin{center}
\includegraphics[width=\columnwidth, angle=0]{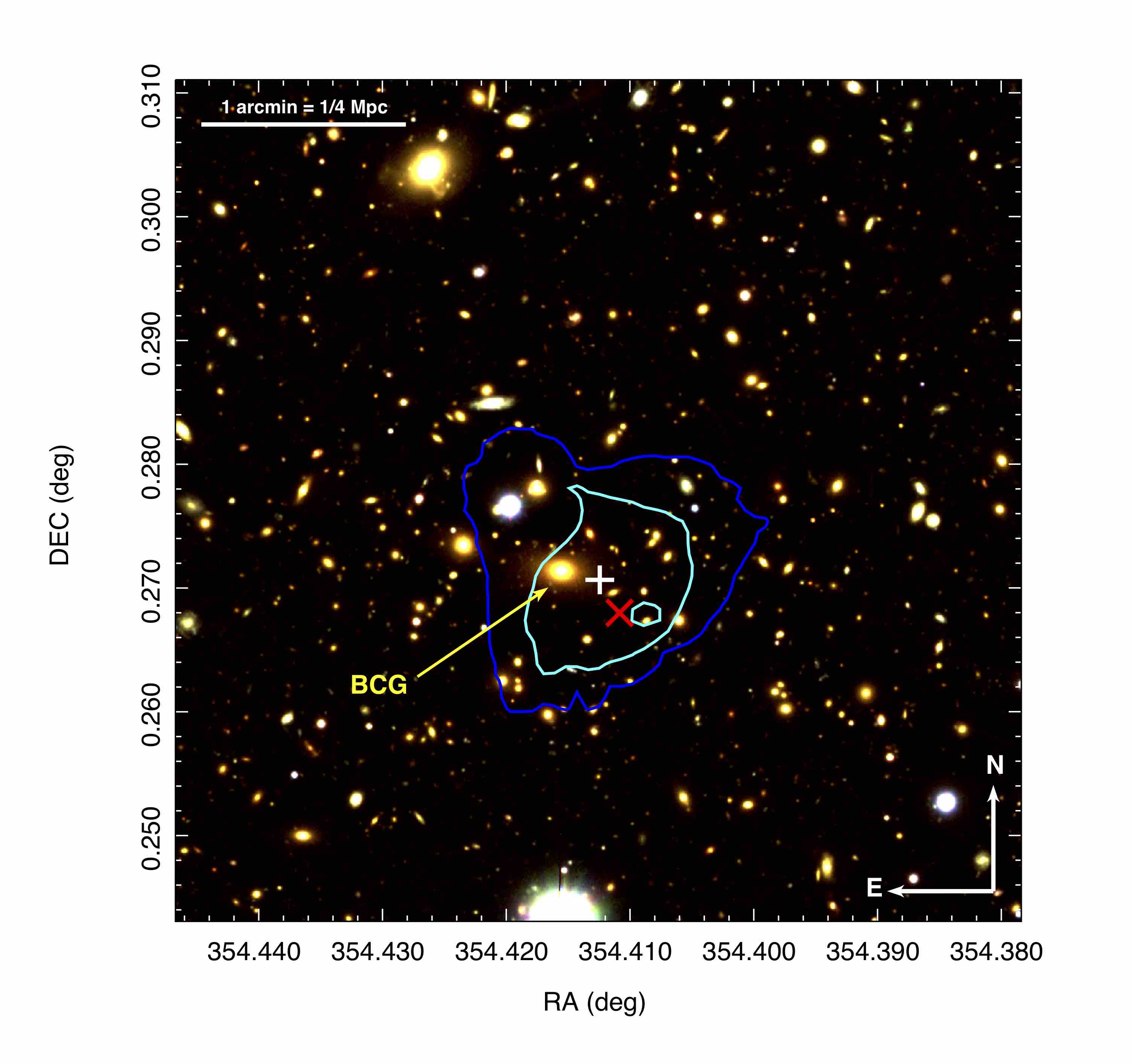}
\caption{Combined $B+V+R_C$ image of the central region of A2631. The cyan and blue contours represent respective 68 and 95 per cent CL of the mass-centre position. The plus sign indicates the face value of our model ($\alpha$: 354.412, $\delta$: 0.271). We also mark the BCG (yellow arrow; $\alpha$: 354.416, $\delta$: 0.271) and the X-ray peak position (red cross; $\alpha :354.411$, $\delta :0.268$) given by \protect\cite{Ge19}.} 
\label{fig:components}
\end{center}
\end{figure}

An important proxy for the dynamical state is the detection of possible spatial detachments among the cluster components, dark matter, galaxies, and the ICM. In Fig.~\ref{fig:components}, we present a close view of A2631 where we can better compare the BCG position (tracer of galaxies distribution), X-ray clump \citep[tracer of the ICM;][]{Ge19}  and mass-centre location (tracer of dark matter). The position of the BCG, X-ray, and dark matter clumps are all consistent within the 68 per cent CL. The mass centre is $23_{-13}^{+10}$ arcsec ($98_{-54}^{+42}$ kpc) away from the BCG location and $21_{-13}^{+10}$ arcsec ($90_{-55}^{+40}$ kpc) from the X-ray clump. The BCG and the X-ray peak are 21 arcsec (90 kpc) apart, which is comparable with the \cite{Mann12} measurement. Therefore, we conclude that all cluster components are centred at a common position within the uncertainties.

\section{Dynamical analysis}
\label{sec:dynamics}

\subsection{Spectroscopic data}
\label{sec:spec.data}

A2631 was intensively observed by SDSS \citep{Alam15} and  the Hectospec Cluster Survey \citep[HeCS,][]{Rines13}. In fact, a search on a circular region centred on the BCG and with a radius of 18 arcmin revealed 418 galaxies with available spectroscopic redshifts and with correspondence in our photometric catalogue.

The respective spectroscopic members were selected after  application of the $3\sigma$ clipping method \citep{3sigmaclip}. This consists of removal of all galaxies beyond the interval $[\bar{z}-3\sigma_z , \bar{z}+3\sigma_z]$, as they are most probably outliers \citep{Wojtak07}.

\begin{figure}
 \begin{center}
\includegraphics[width=\columnwidth, angle=0]{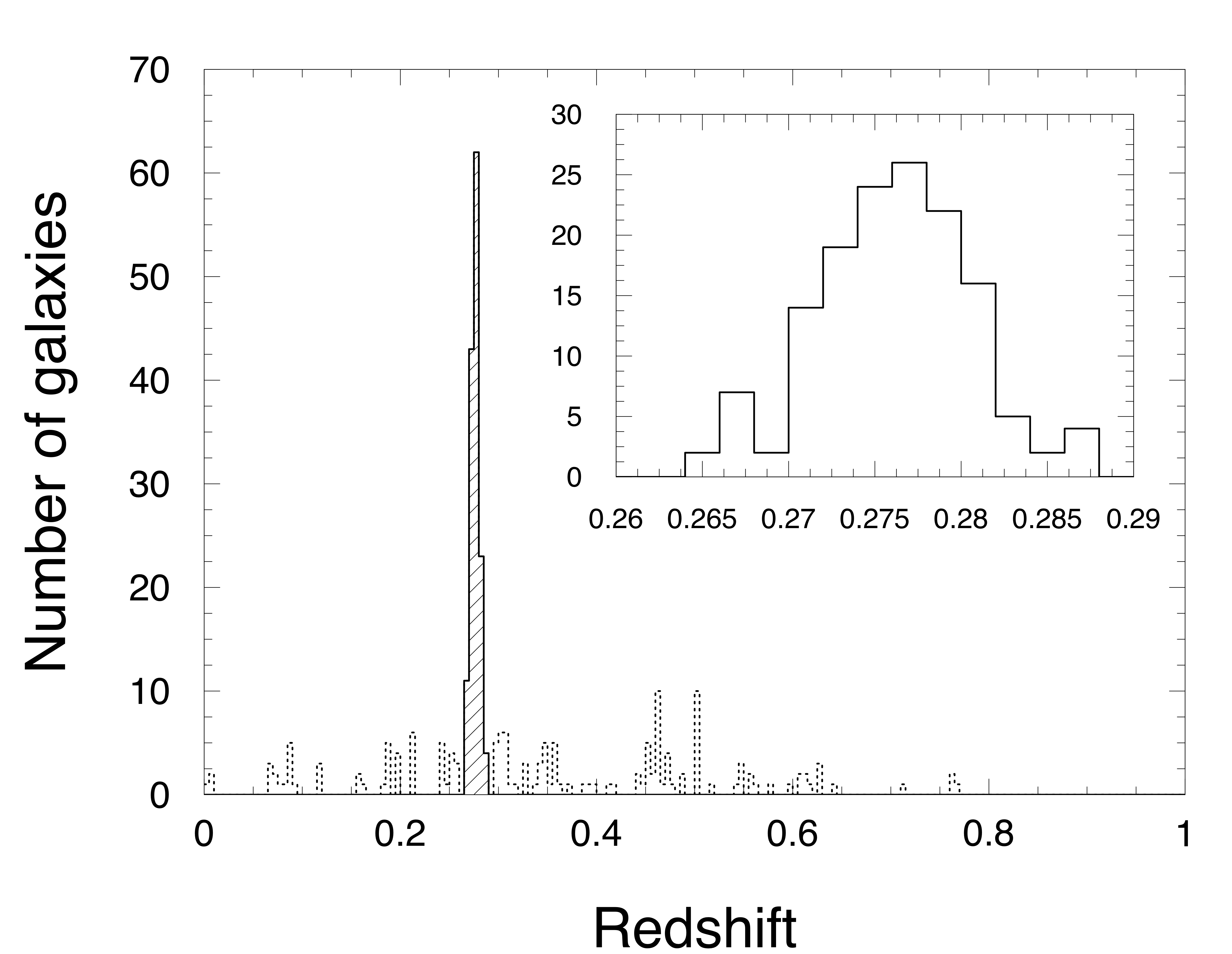}
\caption{Distribution of the redshift catalogue retrieved from the NASA/IPAC Extragalactic Database (NED) inside 18 arcmin around A2631. For aesthetic reasons, the histogram shows the sample with $z\leq0.8$. The selection of spectroscopic cluster members was done through a $3\sigma$-clipping cut (dashed lines).  The inset panel shows a histogram of the 143 selected cluster members with $\bar{z}=0.2762\pm0.0004$ and $\sigma_{v}/(1+\bar{z})=1044_{-56}^{+68}$ km s$^{-1}$.} 
\label{fig:redshift.dist}
\end{center}
\end{figure}

The 143 spectroscopic members of A2631 have $\bar{z}=0.2762\pm 0.0004$ with a standard deviation of $\sigma_{v}/(1+\bar{z})=1044_{-56}^{+68}$ km s$^{-1}$, corresponding to the inset panel in Fig.~\ref{fig:redshift.dist}. According to the Anderson-Darling test \citep{nortest} this sample follows, within the 95 \% CL, a normal distribution (${\rm p-value}=0.15$). This conclusion is also supported by the Hellinger distance estimator \citep{Ribeiro13} within the 92 per cent CL The sample extends up to $\sim 2.5R_{200}$.

\subsection{Search for substructures}
\label{sec:subs}

Even the Gaussianity of the overall redshift distribution does not necessarily mean that the sample is free of substructures. More subtle examples of them (e.g. infalling groups) can pass unscathed through the $3\sigma$ clipping because they have velocities consistent with those of the general cluster population. To scrutinize the sample, we applied the Dressler-Shectman test \citep[or $\Delta$-test;][]{ds}. This consists in quantifying the deviation of the local\footnote{This means that we consider only the $N_{\rm nb}=\sqrt{N}$ closest neighbours, where $N$ is the total number of galaxies in the sample.} systemic velocity and dispersion with those of the overall structure, 
\begin{equation}
 \delta_i=\left \{ \left ( \frac{N_{\rm nb}+1}{\sigma^2} \right ) [(\bar{v}_{l}-\bar{v})^2+(\sigma_l - \sigma)^2] \right \}^{1/2}
 \label{eq:ds}
\end{equation}
and computing the statistic
\begin{equation}
\Delta=\sum_{i=1}^{N} \delta_i.
 \label{eq:sumds}
\end{equation}
It is expected that $\Delta > N$ \citep{ds} with a ${\rm p-value}< 0.01$ \citep{hou12} for substructured clusters. We found $\Delta=176$ with a ${\rm p-value}=0.15$ (95 \% CL) pointing to an absence of substructures in A2631.

The search for substructures based on multidimensional normal mixture modelling was also fruitless. We applied the {\sc R}-based package {\sc MCLUST} \citep{mclust,Lourenco20} in 1D ($z$), 2D ($\alpha$ and $\delta$) and 3D ($\alpha$, $\delta$ and $z$) modes for the spectroscopic members. All of them returned a single group as the best solution, with $\Delta{\rm BIC}\geq10$ in relation to the second-best model with two groups. This means that, according to the dynamical view, A2631 is undoubtedly a unimodal structure. The results remain consistent when we consider smaller radii ($R_{500}$ and $R_{200}$).

\subsection{Dynamical mass}
\label{sec:dyn.mass}

The gravitational potential of the cluster drives the dynamics of its galaxy content. So, we can get the inverse path and estimate the cluster mass from the one-dimensional velocity dispersion since this is an easily measurable observable. In the scope of a virialized system, the virial theorem predicts a theoretical scaling relation in the form $\sigma_{\rm 1D}\propto M_{200}^\alpha$ with $\alpha=1/3$.  This scaling relation has been a matter of intense study from the point of view of computational simulations in order to understand the behaviour of ``real'' systems and then provide a reliable way to determine the cluster mass. 

We can rewrite the scaling relation within a radius $R = R_{200}$ in a more functional form \citep[e.g.][]{biviano06,Evrard08, Munari13},

\begin{equation}
\frac{\sigma}{\rm km \ s^{-1}} = A_{\rm 1D}\left[  \frac{h(z)\ M_{200}}{10^{15} \ {\rm M}_\odot} \right]^\alpha
\label{eq:M.sigma}    
\end{equation}
being $\sigma=\sigma_{\rm 3D}/\sqrt{3}$\footnote{Hereafter, $\sigma=\sigma_{\rm 1D}=\sigma_v/(1+\bar{z})$} and the constants $A_{\rm 1D}$ and $\alpha$ to be determined. Based on realistic baryon (including cooling, star formation and AGN feedback) plus dark matter simulations, \cite{Munari13} suggest $A_{\rm 1D}=1177\pm4.2$ km s$^{-1}$ and $\alpha=0.364\pm0.0021$.

\begin{figure*}
 \begin{center}
\includegraphics[width=\textwidth, angle=0]{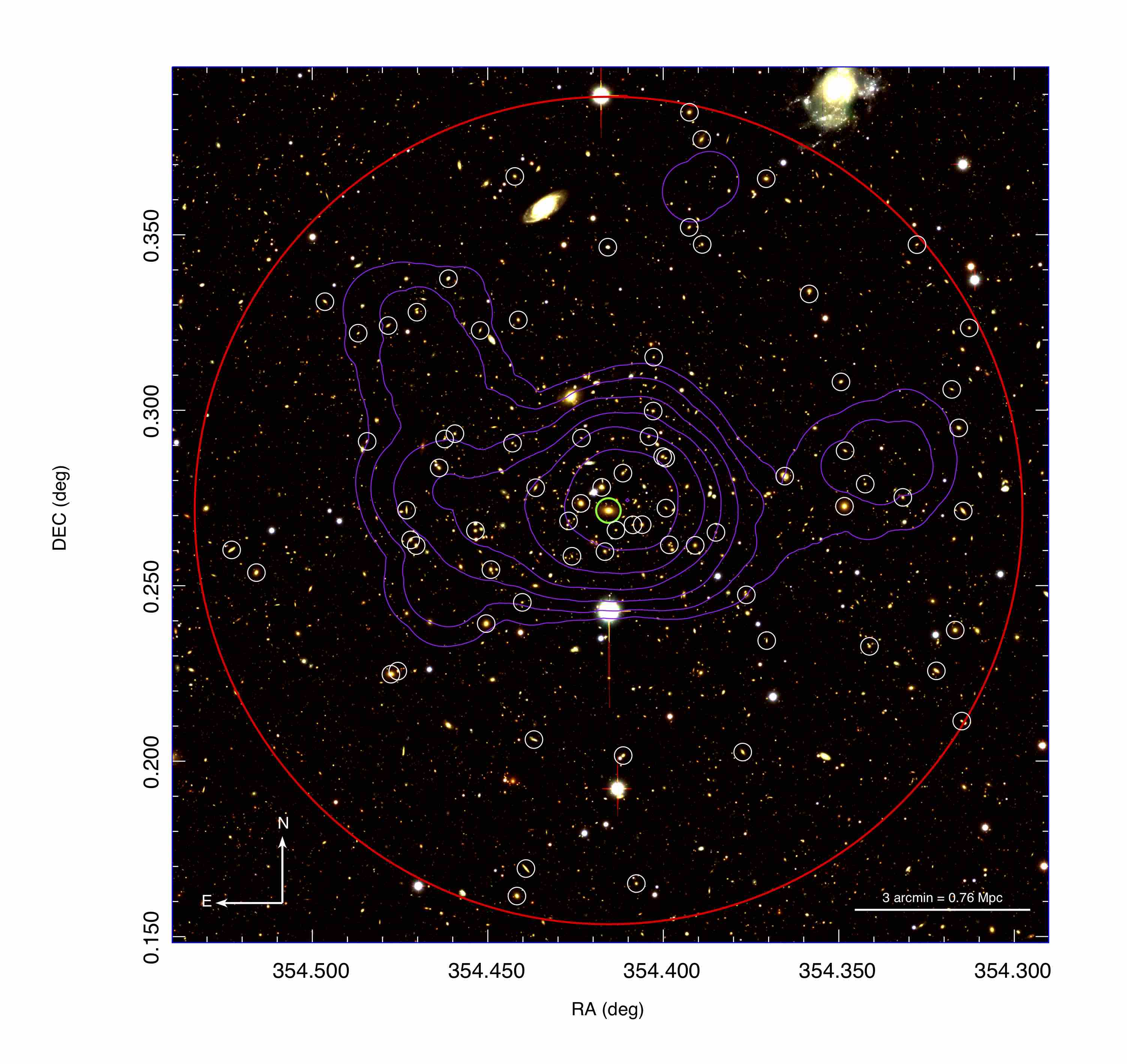} 
\caption{Composite $B+V+R_C$ image of A2631 showing the spectroscopic members inside a radius equal to $R_{200}$ (red circle). The BCG is highlighted with a green circle. We computed the line-of-sight velocity dispersion of these 75 members (small circles) to estimate the cluster dynamic mass $M_{200}^{\rm dy}$. The purple contours show the smoothed spatial distribution.} 
\label{fig:spec.members}
\end{center}
\end{figure*}

Firstly, we should estimate the unbiased velocity dispersion of cluster members. The ``biased'' standard deviation $\sigma (N_{\rm gal})$  is obtained from $N_{\rm gal}= 75$ galaxies inside the projected radius equal to $R_{200}$ (see Fig.~\ref{fig:spec.members}).  We can correct this for the statistical bias induced by the finite number of galaxies as follows \citep{Ferragamo20}:

\begin{equation}
    \sigma^\prime=\sigma(N_{\rm gal}) \left\{ 1+ \left[\left(\frac{D}{N_{\rm gal}-1}\right)^\beta + B \right] \right \} 
\label{eq:sigma.umb}    
\end{equation}
with $D=1/4$, $B=-0.0016\pm0.0005$, and $\beta=1$.

Now, we should correct $\sigma^\prime$ for the bias induced for three main physical effects: $(1)$ the aperture radius where $\sigma (N_{\rm gal})$ is measured, $(2)$ the selected fraction of massive galaxies, and $(3)$ contamination by interlopers. In recent work, \cite{Ferragamo20} presented a comprehensive study regarding statistical properties of velocity dispersion and mass estimators based on simulated galaxy cluster data. The authors suggest a set of multiplicative correction factors to turn $\sigma^\prime$, in fact, into an unbiased estimator. We adopted $f_1=0.998\pm0.001$, since we are working with members enclosed within projected $R_{200}$, $f_2=0.99\pm0.01$, which corresponds to a fraction of 50\%--100\% of the massive galaxies in the cluster present in the sample and $f_3=1.05\pm0.01$ based on the assumption that the 3$\sigma$ clipping selected sample is contaminated by $\sim5\%$ of interlopers \citep{Wojtak07}. After these procedures, we found $\sigma^{\prime\prime}=1120_{-81}^{+104}$ km s$^{-1}$.

Due to the non-linearity of the $\sigma-M_{200}$ relation (Equation~\ref{eq:M.sigma}), even considering an unbiased $\sigma^{\prime\prime}$ we will find dependence on $N_{\rm gal}$ in a  biased mass estimation, especially for the low regime \citep[$N_{\rm gal}\leq 75$;][]{Ferragamo20}. So, we should correct for

\begin{equation}
M^\prime_{200}=M_{200}(\sigma^{\prime\prime}) \left [ \frac{1-E^\prime \alpha}{(E^\prime \alpha)^2(N_{\rm gal}-1)^{\gamma^\prime}}+F^\prime\right]^{-1}
\label{eq:mass.umb}    
\end{equation}
with $E^\prime=1.53\pm0.03$, $F^\prime=1$,  $\gamma^\prime=1.11\pm0.04$, and $M_{200}(\sigma^{\prime\prime})$ the ``biased'' mass given by Equation~\ref{eq:M.sigma}.

The dynamical mass of A2631 is $M_{200}^{\rm dy}=12.2\pm3.0\times10^{14}$ M$_\odot$. This corresponds to a face value $\sim 40$ per cent larger than the weak-lensing estimated mass (Table~\ref{tab:masses.dm}).  However, both mass measurements are consistent within the 68 per cent level. We will resume this discussion in the next section.

\section{Discussion}
\label{sec:discussion}

\subsection{Weak-lensing mass of A2631}
\label{sec:disc.wl}

We present here a comprehensive weak-lensing analysis of the massive galaxy cluster A2631 ($\bar{z}=0.2762$) located at the centre of the Saraswati supercluster. In order to provide an unambiguous estimation of the cluster mass, we reconstructed the mass field combining measurements of shape distortion and magnification bias of the background galaxies. This is a powerful tool because it combines the strengths of both observables: whereas the distortion provides a high S/N, the magnification bias is insensitive to the mass-sheet degeneracy. We based our analysis on large field-of-view  ($37 \times 37.5$ arcmin$^2$) multiband ($B, R_C, V$)  Subaru archival images. The source catalogue was carefully selected in the colour-colour space, aiming for a final catalogue as pure as possible (i.e. with minimal contamination by cluster/foreground galaxies). The purity of the source catalogue is essential to preserve the lens-induced signal and then obtain a credible determination of the original mass field.

Our MCMC-based model confirmed that A2631 is a very massive cluster having $M_{200}^{\rm wl}=8.7_{-2.9}^{+2.5}\times 10^{14}$ M$_\odot$. Indeed, according to the Schechter cluster mass function for nearby clusters presented by \cite{Girardi98b}, A2631 is among the $\sim4$ per cent of most massive clusters with masses above $M^\star=2.6_{-0.6}^{+0.8}\times 10^{14}$ M$_\odot$. The cluster mass corresponds to a halo concentration of $c_{200}=3.2\pm0.1$ according to the $M_{200}- c_{200}$ relation proposed by \cite{duffy08}.  We also find that the BCG, X-ray emission peak, and  mass centre are nearly concentric, all being coincident within the 68 per cent CL. The combination of distortion and magnification bias decreased the uncertainties on the masses to $\sim78$ per cent of the distortion-only estimation.

\begin{figure*}
\begin{center}
\includegraphics[width=\textwidth, angle=0]{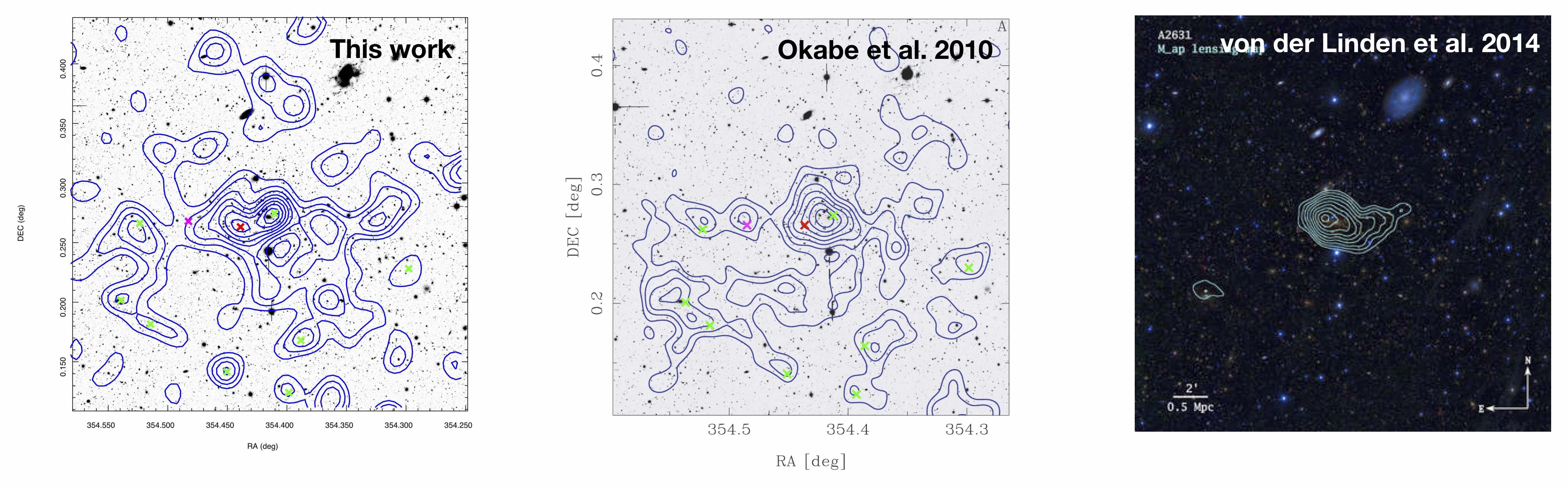}
\caption{Comparison of the mass maps of A2631 available in the literature. We resized our convergence map to match \protect\cite{Okabe10} ($20 \times 20$ arcmin$^2$) and we start the contours at the $3\sigma_\kappa$ level. The aperture mass map of \protect\cite{vonderLinden14} is somewhat larger ($37 \times 37$ arcmin$^2$). Halo B in Fig.~\ref{fig:mass.map} is marked with a red cross.  The green crosses show the common structures found in both maps whereas the magenta one shows a structure detected only by \protect\cite{Okabe10}.} 
\label{fig:mass.maps.comp}
\end{center}
\end{figure*}

A visual inspection of the convergence map (Fig.~\ref{fig:mass.map}) can induce the reader to suppose the presence of a companion structure east of A2631. However, this scenario is not supported  by either the luminosity-weighted spatial map of the red-sequence galaxies (Fig.~\ref{fig:field}) or the projected distribution of the spectroscopic members (Fig.~\ref{fig:spec.members}). Lastly, our mass modelling statistics completely refute any evidence of additional halos in the field. So, what would be the nature of those apparent mass clumps? To answer this question, in  Fig.~\ref{fig:mass.maps.comp} we compare our convergence map with  \cite{Okabe10} and the mass aperture map from \cite{vonderLinden14}. The three maps show a dominant mass clump surrounding the BCG. However, a detailed comparison can only be done with  \cite{Okabe10} since \cite{vonderLinden14} shows only the highest-density regions.

Both maps show several structures across the field with most of them found in common (green crosses). Although suggesting a little elongation in the E--W direction when we consider the fewer significative contours, \cite{Okabe10} did not find a relevant mass neighbour to A2631. So, we conjecture that it arises in our map from the association of small background halos ($\lesssim 10^{13}$ M$_\odot$) since they dominate the population in the field \citep{Yang13, Liu16, Wei18}.

\begin{figure}
\begin{center}
\includegraphics[width=\columnwidth, angle=0]{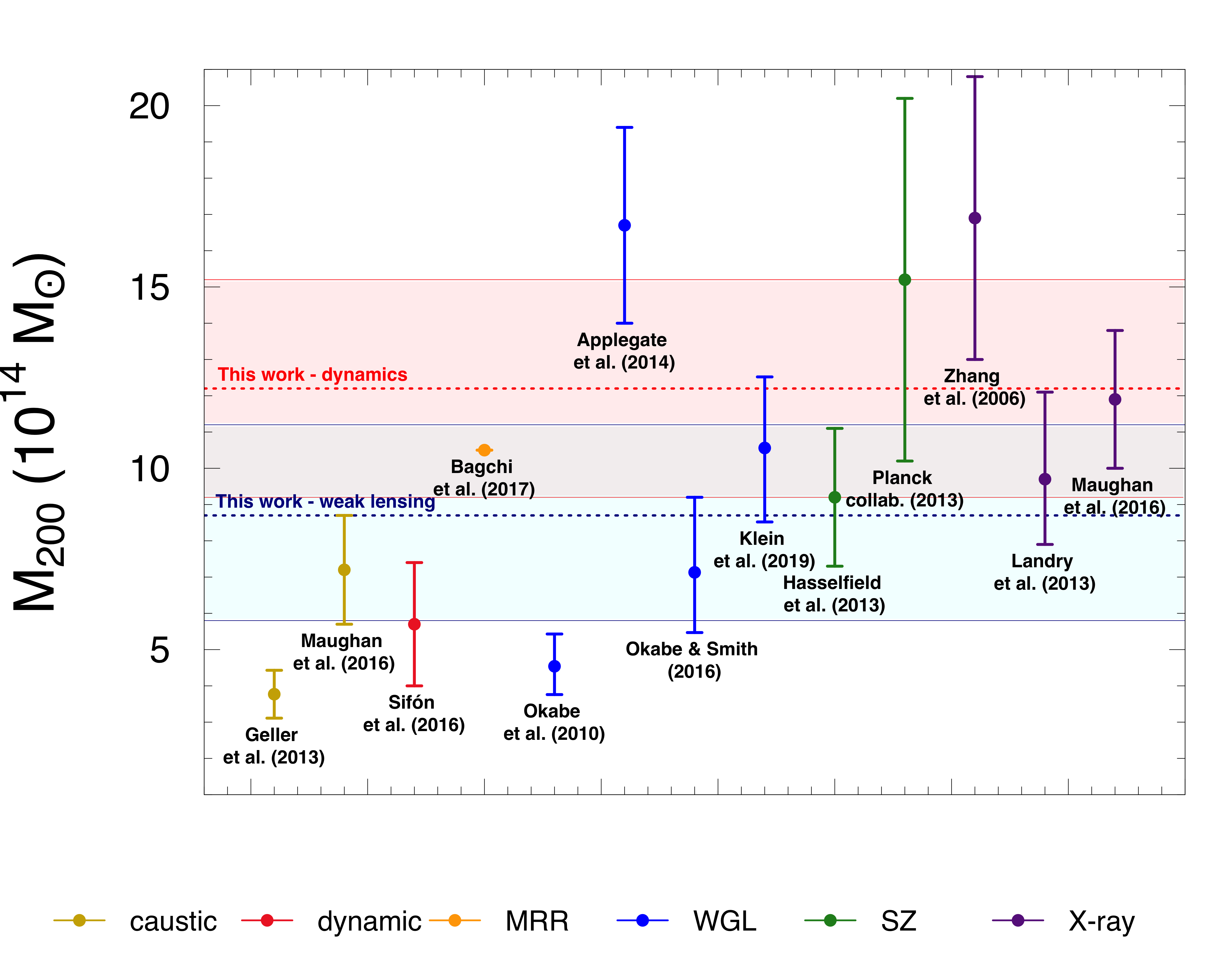} 
\caption{Comparison of our mass estimates (weak-lensing and dynamic) with those found in literature as presented in Table~\ref{tab:mass.comp}. The shadows represent the 68 per cent CL.} 
\label{fig:mass.comp}
\end{center}
\end{figure}

Now, we will compare quantitatively our results with previous weak-lensing-based works. They are presented in Fig.~\ref{fig:mass.comp} by blue points. Our fiducial mass estimation shows, within the 68 per cent CL, a very good match with \cite{Okabe16} and \cite{Klein19}. However, within the same confidence level, the values of \cite{Okabe10} and \cite{vonderLinden14} are inconsistent with ours.

Among the available weak-lensing mass estimates, \cite{Okabe10} present the lowest value. Checking the work we found that the authors had only two bands ($R_C$ and $V$) to select the sources. Despite their care, this could introduce a high degree of contamination leading to a biased increasing of the sample size. In fact, their final sample had a very high density of 31 galaxies per arcmin$^{2}$ against $13.5$ galaxies per arcmin$^{2}$ in this work. For the sake of comparison, in \cite{Monteiro-Oliveira17a} with an exposure time $2.5\times$ higher, we found a source density of $\sim24$ galaxies per arcmin$^{2}$ in the $z'$ band for a cluster located at a similar redshift. Another important point is that, unlike our analysis, the authors left the halo concentration as a free parameter. To check the possible impact on the concentration in our results, we considered an additional model where we kept the centre fixed at the corresponding mass peak and modelled  $M_{200}$ and $c_{200}$.  We found $M_{200_c}=7.0_{-2.5}^{+2.8}\times 10^{14}$ M$_\odot$, a value still much higher than those provided by \cite{Okabe10}. The concentration $c_{200}=4.6_{-2.0}^{+2.9}$ is in agreement with those provided by \cite{duffy08} but is lower than \cite{Okabe10}\footnote{In their work, \cite{Okabe10} found $c_{\rm vir}=7.84_{-2.28}^{+3.54}$ corresponding to $M_{\rm vir}=5.24_{-0.98}^{+1.15}\times 10^{14}$} M$_\odot$. It is known that the measurement of the concentration is more sensitive to the cluster centric region. At the same time, the source contamination becomes more dramatic in this region. In this sense, we speculate that the contamination of the source sample in \cite{Okabe10} could be responsible for their low mass estimation, in spite of the mass maps being very similar as can be seen in Fig.~\ref{fig:mass.maps.comp}.

Within the scope of the WtG project, \cite{Applegate14} determined the mass of A2631, but their value corresponds to the largest available estimate in the literature. For comparison,  their modelled mass corresponds to $\sim 2\times$ our face value. In addition to us, several other authors \citep{Hoekstra15,Umetsu16,Okabe16} have also found a tension when comparing their masses with those from WtG. A general comparison shows that the WtG masses are biased higher than other mass surveys \citep[CCCP, CLASH and LoCuSS;][]{Pratt19}, in agreement with our conclusion. One reason for this, according to \cite{Okabe16}, could be the use of a pre-fixed $c_{200}=4$ and/or the construction of the WtG source catalogue which could have introduced a high degree of contamination by cluster/foreground galaxies.

After these considerations, we conclude that the mass estimate of A2631 presented in this work is comparable with those given by \cite{Okabe16} and \cite{Klein19} and note that the estimations of \cite{Okabe10} and \cite{Applegate14} can be considered as outliers. We reinforce that the main strength of the present work is to provide a mass estimate based on two different observables (shear and magnification) leading to a confident constraint for the cluster mass and providing more accurate data for cosmological applications based on galaxy cluster masses.

\subsection{Dynamic state of A2631}
\label{sec:disc.dy}

With the catalogue of spectroscopic members of A2631, we investigated the cluster structure from the dynamical point of view. Undisturbed clusters are expected to have a Gaussian distribution of member galaxy redshifts whereas non-Gaussian distributions are a tracer of disturbed systems \citep{Ribeiro13}. However, some highly disturbed clusters  \citep{Einasto15, Monteiro-Oliveira17a, Monteiro-Oliveira17b, Monteiro-Oliveira18, Monteiro-Oliveira20} have  Gaussian distributions because most of the three-dimensional movement's component is parallel to the plane of the sky.

We found that our sample follows, with a 95 per cent CL, a Gaussian distribution. A suite of substructuring tests has shown a scenario that supports that seen in the mass distribution, i.e., a unimodal halo. There are no signals of substructure up to a projected radius $\sim 2.5R_{200}$ according to the multidimensional normal mixture modelling employed in 1D, 2D and 3D coordinates.

We estimate a dynamical mass of $M_{200}^{\rm dy}=12.2\pm3.0\times 10^{14}$ M$_\odot$. Although 40 per cent larger than $M_{200}^{\rm wl}$, both estimations are consistent within 1$\sigma$, as we can see from Fig.~\ref{fig:mass.comp}.

We can obtain additional insights about the disturbed state of galaxy clusters by a comparison between the measured velocity dispersion with that expected before a possible cluster interaction process. This assumption relies on the fact that the velocity dispersion is boosted during the merger process \citep[e.g.][]{pinkney,Takizawa10, Monteiro-Oliveira20}. Considering a minor merger \cite{martel14}, we can disregard as a first approximation the mass of the infalling subcluster. Then, we can estimate the pre-merger velocity dispersion $\sigma_{\rm pre}$ with Equation~\ref{eq:M.sigma}. We found $\sigma_{\rm pre}/(1+\bar{z})=1030_{-92}^{+135}$  km s$^{-1}$, leading to a boost factor $f \equiv \sigma_{\rm obs} / \sigma_{\rm pre}=1.1_{-0.2}^{+0.1}$. Despite a probability of 28 per cent that $f$ is lower than unity, the boost factor is consistant with  unity, meaning that there is no significant signature of the merger process on the galaxy dynamics.

A comparison of the mass estimates at different wavelength provides of another useful piece of information to constrain the cluster dynamical state \citep[e.g.][]{cypriano04, Soja18}. In disturbed clusters, the hydrostatic equilibrium is violated changing the scaling relations \citep[e.g. $M-T_X$ or $M-L_X$;][]{Andrade-Santos12} and biasing the X-ray and SZ-based mass estimation. Overall, according to Fig.~\ref{fig:mass.comp} all mass estimations based on X-ray and Sunyaev-Zel'dovich data have a good match with our $M_{200}^{\rm wl}$ within the 1$\sigma$ level. An exception is the estimate of \cite{Zhang06}, despite its large error bars.

Based on the above statements, we can draw a general picture of the A2631 dynamical state. Considering only the dark matter distribution and galaxy dynamics, it is possible to affirm that the scenario found for A2631 is coherent with a non-disturbed system. The same picture is drawn after comparison with mass estimations at other wavelengths (X-ray and SZ). However, \cite{Finoguenov05, Zhang06, Mann12, Marrone12} found strong evidence of disturbance in the innermost ($r \lesssim R_{500}$) region of ICM. How can we reconcile these two contrasting descriptions?

The cluster components interact with themselves differently during the merger process. This fact is quantified in terms of the self-interacting cross-section  \citep[$\sigma_{\rm si}$; e.g.][]{markevitch04}. While galaxies have $\sigma_{\rm si}=0$ and dark matter has $\sigma_{\rm si}/m \lesssim 2$ \citep[e.g.][]{ Wittman18a, Drlica-Wagner19, Sagunski20}, the fluid nature of the gas makes its corresponding $\sigma_{\rm si}$ naturally larger.  This means that the gas will suffer the most dramatic events during a cluster merger \cite[e.g.][]{markevitch_viki07}. In some extreme cases, the gas is stripped from its host halo \citep[e.g.][]{harvey15, Doubrawa20}. As a consequence, the ICM can retain a memory of the merger process for greater intervals of time in comparison with dark matter and galaxies, and is therefore, the last component to reach a state of equilibrium. 

We conclude this work stating that the scenario that reconciles the findings of our dynamical analysis with the previous observation of the ICM is that described by A2631  experiencing a late stage of the cluster merger process.

\section{Concluding remarks}
\label{sec:summary}

A2631 is a unimodal cluster as indicated by its red-sequence and dark matter projected spatial distributions. Through the use of the combined weak gravitational lensing signal from shear and magnification, we constrained the mass of A2631 as $M_{200}^{\rm wl}=8.7_{-2.9}^{+2.5}\times 10^{14}$ M$_\odot$. We observed a spatial coincidence between the positions of the BCG, the peak of X-ray emission, and the centre of the dark matter distribution. The radial velocities of the spectroscopic members show no evidence of the presence of substructures in the cluster. We estimated the dynamical mass of A2631 as $M_{200}^{\rm dy}=12.2\pm3.0\times 10^{14}$ M$_\odot$, which is comparable (within 1$\sigma$) with the weak-lensing mass. Finally, we concluded that the scenario found in A2631 is consistent with a galaxy cluster in a late merger stage.

\section*{Acknowledgements}
\addcontentsline{toc}{section}{Acknowledgements}

RMO thanks Eduardo S. Cypriano for kindly sharing his computational resources that made this work possible. ALBR is grateful for the support of CNPq, grant 311932/2017-7.

This work is based on data collected by the Subaru Telescope and obtained from SMOKA, which is operated by the Astronomy Data Center, National Astronomical Observatory of Japan.

We made use of the NASA/IPAC Extragalactic Database, which is operated by the Jet Propulsion Laboratory, California Institute of Technology, under contract with NASA.

\section*{Data Availability}

Raw data are currently available from the respective sources. The reduced data underlying this article will be shared on reasonable request to the corresponding author.




\bibliographystyle{mnras}
\bibliography{monteiro-oliveira_library}

\begin{thebibliography}{}
\makeatletter
\relax
\def\mn@urlcharsother{\let\do\@makeother \do\$\do\&\do\#\do\^\do\_\do\%\do\~}
\def\mn@doi{\begingroup\mn@urlcharsother \@ifnextchar [ {\mn@doi@}
  {\mn@doi@[]}}
\def\mn@doi@[#1]#2{\def\@tempa{#1}\ifx\@tempa\@empty \href
  {http://dx.doi.org/#2} {doi:#2}\else \href {http://dx.doi.org/#2} {#1}\fi
  \endgroup}
\def\mn@eprint#1#2{\mn@eprint@#1:#2::\@nil}
\def\mn@eprint@arXiv#1{\href {http://arxiv.org/abs/#1} {{\tt arXiv:#1}}}
\def\mn@eprint@dblp#1{\href {http://dblp.uni-trier.de/rec/bibtex/#1.xml}
  {dblp:#1}}
\def\mn@eprint@#1:#2:#3:#4\@nil{\def\@tempa {#1}\def\@tempb {#2}\def\@tempc
  {#3}\ifx \@tempc \@empty \let \@tempc \@tempb \let \@tempb \@tempa \fi \ifx
  \@tempb \@empty \def\@tempb {arXiv}\fi \@ifundefined
  {mn@eprint@\@tempb}{\@tempb:\@tempc}{\expandafter \expandafter \csname
  mn@eprint@\@tempb\endcsname \expandafter{\@tempc}}}

\bibitem[\protect\citeauthoryear{{Alam} et~al.,}{{Alam} et~al.}{2015}]{Alam15}
{Alam} S.,  et~al., 2015, \mn@doi [The Astrophysical Journal Supplement Series]
  {10.1088/0067-0049/219/1/12}, \href
  {https://ui.adsabs.harvard.edu/#abs/2015ApJS..219...12A} {219, 12}

\bibitem[\protect\citeauthoryear{{Andrade-Santos}, {Lima Neto}  \&
  {Lagan{\'a}}}{{Andrade-Santos} et~al.}{2012}]{Andrade-Santos12}
{Andrade-Santos} F.,  {Lima Neto} G.~B.,   {Lagan{\'a}} T.~F.,  2012, \mn@doi
  [\apj] {10.1088/0004-637X/746/2/139}, \href
  {https://ui.adsabs.harvard.edu/abs/2012ApJ...746..139A} {746, 139}

\bibitem[\protect\citeauthoryear{{Applegate} et~al.,}{{Applegate}
  et~al.}{2014}]{Applegate14}
{Applegate} D.~E.,  et~al., 2014, \mn@doi [\mnras] {10.1093/mnras/stt2129},
  \href {https://ui.adsabs.harvard.edu/abs/2014MNRAS.439...48A} {439, 48}

\bibitem[\protect\citeauthoryear{{Bagchi}, {Sankhyayan}, {Sarkar},
  {Raychaudhury}, {Jacob}  \& {Dabhade}}{{Bagchi} et~al.}{2017}]{Bagchi17}
{Bagchi} J.,  {Sankhyayan} S.,  {Sarkar} P.,  {Raychaudhury} S.,  {Jacob} J.,
  {Dabhade} P.,  2017, \mn@doi [\apj] {10.3847/1538-4357/aa7949}, \href
  {https://ui.adsabs.harvard.edu/abs/2017ApJ...844...25B} {844, 25}

\bibitem[\protect\citeauthoryear{{Bertin} \& {Arnouts}}{{Bertin} \&
  {Arnouts}}{1996}]{sextractor}
{Bertin} E.,  {Arnouts} S.,  1996, \aaps, \href
  {http://adsabs.harvard.edu/abs/1996A%26AS..117..393B} {117, 393}

\bibitem[\protect\citeauthoryear{{Biviano}, {Murante}, {Borgani}, {Diaferio},
  {Dolag}  \& {Girardi}}{{Biviano} et~al.}{2006}]{biviano06}
{Biviano} A.,  {Murante} G.,  {Borgani} S.,  {Diaferio} A.,  {Dolag} K.,
  {Girardi} M.,  2006, \mn@doi [\aap] {10.1051/0004-6361:20064918}, \href
  {http://adsabs.harvard.edu/abs/2006A%26A...456...23B} {456, 23}

\bibitem[\protect\citeauthoryear{{Bridle}, {Hobson}, {Lasenby}  \&
  {Saunders}}{{Bridle} et~al.}{1998}]{im2shape}
{Bridle} S.~L.,  {Hobson} M.~P.,  {Lasenby} A.~N.,   {Saunders} R.,  1998,
  \mn@doi [\mnras] {10.1046/j.1365-8711.1998.01877.x}, \href
  {http://adsabs.harvard.edu/abs/1998MNRAS.299..895B} {299, 895}

\bibitem[\protect\citeauthoryear{{Capak} et~al.,}{{Capak}
  et~al.}{2007}]{capak07}
{Capak} P.,  et~al., 2007, \mn@doi [\apjs] {10.1086/519081}, \href
  {http://adsabs.harvard.edu/abs/2007ApJS..172...99C} {172, 99}

\bibitem[\protect\citeauthoryear{{Cypriano}, {Sodr{\'e}}, {Kneib}  \&
  {Campusano}}{{Cypriano} et~al.}{2004}]{cypriano04}
{Cypriano} E.~S.,  {Sodr{\'e}} Jr. L.,  {Kneib} J.-P.,   {Campusano} L.~E.,
  2004, \mn@doi [\apj] {10.1086/422896}, \href
  {http://adsabs.harvard.edu/abs/2004ApJ...613...95C} {613, 95}

\bibitem[\protect\citeauthoryear{{Doubrawa}, {Machado}, {Lagan{\'a}}, {Lima
  Neto}, {Monteiro-Oliveira}  \& {Cypriano}}{{Doubrawa}
  et~al.}{2020}]{Doubrawa20}
{Doubrawa} L.,  {Machado} R.~E.~G.,  {Lagan{\'a}} T.~F.,  {Lima Neto} G.~B.,
  {Monteiro-Oliveira} R.,   {Cypriano} E.~S.,  2020, \mn@doi [\mnras]
  {10.1093/mnras/staa1051}, \href
  {https://ui.adsabs.harvard.edu/abs/2020MNRAS.495.2022D} {495, 2022}

\bibitem[\protect\citeauthoryear{{Dressler}}{{Dressler}}{1980}]{dressler80}
{Dressler} A.,  1980, \mn@doi [\apj] {10.1086/157753}, \href
  {http://adsabs.harvard.edu/abs/1980ApJ...236..351D} {236, 351}

\bibitem[\protect\citeauthoryear{{Dressler} \& {Shectman}}{{Dressler} \&
  {Shectman}}{1988}]{ds}
{Dressler} A.,  {Shectman} S.~A.,  1988, \mn@doi [\aj] {10.1086/114694}, \href
  {http://adsabs.harvard.edu/abs/1988AJ.....95..985D} {95, 985}

\bibitem[\protect\citeauthoryear{{Drlica-Wagner} et~al.,}{{Drlica-Wagner}
  et~al.}{2019}]{Drlica-Wagner19}
{Drlica-Wagner} A.,  et~al., 2019, arXiv e-prints, \href
  {https://ui.adsabs.harvard.edu/abs/2019arXiv190201055D} {p. arXiv:1902.01055}

\bibitem[\protect\citeauthoryear{{Duffy}, {Schaye}, {Kay}  \& {Dalla
  Vecchia}}{{Duffy} et~al.}{2008}]{duffy08}
{Duffy} A.~R.,  {Schaye} J.,  {Kay} S.~T.,   {Dalla Vecchia} C.,  2008, \mn@doi
  [\mnras] {10.1111/j.1745-3933.2008.00537.x}, \href
  {http://adsabs.harvard.edu/abs/2008MNRAS.390L..64D} {390, L64}

\bibitem[\protect\citeauthoryear{{Einasto} et~al.,}{{Einasto}
  et~al.}{2015}]{Einasto15}
{Einasto} M.,  et~al., 2015, \mn@doi [\aap] {10.1051/0004-6361/201526399},
  \href {https://ui.adsabs.harvard.edu/abs/2015A&A...580A..69E} {580, A69}

\bibitem[\protect\citeauthoryear{{Evrard} et~al.,}{{Evrard}
  et~al.}{2008}]{Evrard08}
{Evrard} A.~E.,  et~al., 2008, \mn@doi [\apj] {10.1086/521616}, \href
  {https://ui.adsabs.harvard.edu/abs/2008ApJ...672..122E} {672, 122}

\bibitem[\protect\citeauthoryear{{Feretti}, {Gioia}  \& {Giovannini}}{{Feretti}
  et~al.}{2002}]{merging_book}
{Feretti} L.,  {Gioia} I.~M.,   {Giovannini} G.,  eds, 2002, {Merging Processes
  in Galaxy Clusters}  Astrophysics and Space Science Library Vol. 272,
  \mn@doi{10.1007/0-306-48096-4.
}

\bibitem[\protect\citeauthoryear{{Ferragamo}, {Rubi{\~n}o-Mart{\'\i}n},
  {Betancort-Rijo}, {Munari}, {Sartoris}  \& {Barrena}}{{Ferragamo}
  et~al.}{2020}]{Ferragamo20}
{Ferragamo} A.,  {Rubi{\~n}o-Mart{\'\i}n} J.~A.,  {Betancort-Rijo} J.,
  {Munari} E.,  {Sartoris} B.,   {Barrena} R.,  2020, \mn@doi [\aap]
  {10.1051/0004-6361/201834837}, \href
  {https://ui.adsabs.harvard.edu/abs/2020A&A...641A..41F} {641, A41}

\bibitem[\protect\citeauthoryear{{Finoguenov}, {B{\"o}hringer}  \&
  {Zhang}}{{Finoguenov} et~al.}{2005}]{Finoguenov05}
{Finoguenov} A.,  {B{\"o}hringer} H.,   {Zhang} Y.~Y.,  2005, \mn@doi [\aap]
  {10.1051/0004-6361:20053306}, \href
  {https://ui.adsabs.harvard.edu/abs/2005A&A...442..827F} {442, 827}

\bibitem[\protect\citeauthoryear{{Fukugita}, {Shimasaku}  \&
  {Ichikawa}}{{Fukugita} et~al.}{1995}]{fukugita}
{Fukugita} M.,  {Shimasaku} K.,   {Ichikawa} T.,  1995, \mn@doi [\pasp]
  {10.1086/133643}, \href {http://adsabs.harvard.edu/abs/1995PASP..107..945F}
  {107, 945}

\bibitem[\protect\citeauthoryear{{Ge}, {Sun}, {Rozo}, {Sehgal}, {Vikhlinin},
  {Forman}, {Jones}  \& {Nagai}}{{Ge} et~al.}{2019}]{Ge19}
{Ge} C.,  {Sun} M.,  {Rozo} E.,  {Sehgal} N.,  {Vikhlinin} A.,  {Forman} W.,
  {Jones} C.,   {Nagai} D.,  2019, \mn@doi [\mnras] {10.1093/mnras/stz088},
  \href {https://ui.adsabs.harvard.edu/abs/2019MNRAS.484.1946G} {484, 1946}

\bibitem[\protect\citeauthoryear{{Geller}, {Diaferio}, {Rines}  \&
  {Serra}}{{Geller} et~al.}{2013}]{Geller13}
{Geller} M.~J.,  {Diaferio} A.,  {Rines} K.~J.,   {Serra} A.~L.,  2013, \mn@doi
  [\apj] {10.1088/0004-637X/764/1/58}, \href
  {https://ui.adsabs.harvard.edu/abs/2013ApJ...764...58G} {764, 58}

\bibitem[\protect\citeauthoryear{{Girardi}, {Borgani}, {Giuricin},
  {Mardirossian}  \& {Mezzetti}}{{Girardi} et~al.}{1998}]{Girardi98b}
{Girardi} M.,  {Borgani} S.,  {Giuricin} G.,  {Mardirossian} F.,   {Mezzetti}
  M.,  1998, \mn@doi [\apj] {10.1086/306252}, \href
  {https://ui.adsabs.harvard.edu/abs/1998ApJ...506...45G} {506, 45}

\bibitem[\protect\citeauthoryear{Gross \& Ligges}{Gross \&
  Ligges}{2012}]{nortest}
Gross J.,  Ligges U.,  2012, nortest: Tests for Normality.
\url {http://CRAN.R-project.org/package=nortest}

\bibitem[\protect\citeauthoryear{{Haines}, {Smith}, {Egami}, {Okabe}, {Takada},
  {Ellis}, {Moran}  \& {Umetsu}}{{Haines} et~al.}{2009}]{Haines09}
{Haines} C.~P.,  {Smith} G.~P.,  {Egami} E.,  {Okabe} N.,  {Takada} M.,
  {Ellis} R.~S.,  {Moran} S.~M.,   {Umetsu} K.,  2009, \mn@doi [\mnras]
  {10.1111/j.1365-2966.2009.14823.x}, \href
  {http://adsabs.harvard.edu/abs/2009MNRAS.396.1297H} {396, 1297}

\bibitem[\protect\citeauthoryear{{Harvey}, {Massey}, {Kitching}, {Taylor}  \&
  {Tittley}}{{Harvey} et~al.}{2015}]{harvey15}
{Harvey} D.,  {Massey} R.,  {Kitching} T.,  {Taylor} A.,   {Tittley} E.,  2015,
  \mn@doi [Science] {10.1126/science.1261381}, \href
  {http://adsabs.harvard.edu/abs/2015Sci...347.1462H} {347, 1462}

\bibitem[\protect\citeauthoryear{{Hasselfield} et~al.,}{{Hasselfield}
  et~al.}{2013}]{Hasselfield13}
{Hasselfield} M.,  et~al., 2013, \mn@doi [\jcap]
  {10.1088/1475-7516/2013/07/008}, \href
  {https://ui.adsabs.harvard.edu/abs/2013JCAP...07..008H} {2013, 008}

\bibitem[\protect\citeauthoryear{{Hoekstra}, {Herbonnet}, {Muzzin}, {Babul},
  {Mahdavi}, {Viola}  \& {Cacciato}}{{Hoekstra} et~al.}{2015}]{Hoekstra15}
{Hoekstra} H.,  {Herbonnet} R.,  {Muzzin} A.,  {Babul} A.,  {Mahdavi} A.,
  {Viola} M.,   {Cacciato} M.,  2015, \mn@doi [\mnras] {10.1093/mnras/stv275},
  \href {https://ui.adsabs.harvard.edu/abs/2015MNRAS.449..685H} {449, 685}

\bibitem[\protect\citeauthoryear{{Hou} et~al.,}{{Hou} et~al.}{2012}]{hou12}
{Hou} A.,  et~al., 2012, \mn@doi [\mnras] {10.1111/j.1365-2966.2012.20586.x},
  \href {http://adsabs.harvard.edu/abs/2012MNRAS.421.3594H} {421, 3594}

\bibitem[\protect\citeauthoryear{{Ilbert} et~al.,}{{Ilbert}
  et~al.}{2009}]{Ilbert09}
{Ilbert} O.,  et~al., 2009, \mn@doi [\apj] {10.1088/0004-637X/690/2/1236},
  \href {https://ui.adsabs.harvard.edu/abs/2009ApJ...690.1236I} {690, 1236}

\bibitem[\protect\citeauthoryear{{Kass} \& {Raftery}}{{Kass} \&
  {Raftery}}{1995}]{kass95}
{Kass} R.~E.,  {Raftery} A.~E.,  1995, Journal of the American Statistical
  Association, 90, 773

\bibitem[\protect\citeauthoryear{{Klein}, {Israel}, {Nagarajan}, {Bertoldi},
  {Pacaud}, {Lee}, {Sommer}  \& {Basu}}{{Klein} et~al.}{2019}]{Klein19}
{Klein} M.,  {Israel} H.,  {Nagarajan} A.,  {Bertoldi} F.,  {Pacaud} F.,  {Lee}
  A.~T.,  {Sommer} M.,   {Basu} K.,  2019, \mn@doi [\mnras]
  {10.1093/mnras/stz1491}, \href
  {https://ui.adsabs.harvard.edu/abs/2019MNRAS.488.1704K} {488, 1704}

\bibitem[\protect\citeauthoryear{{Knowles} et~al.,}{{Knowles}
  et~al.}{2019}]{Knowles19}
{Knowles} K.,  et~al., 2019, \mn@doi [\mnras] {10.1093/mnras/stz823}, \href
  {https://ui.adsabs.harvard.edu/abs/2019MNRAS.486.1332K} {486, 1332}

\bibitem[\protect\citeauthoryear{{Kravtsov} \& {Borgani}}{{Kravtsov} \&
  {Borgani}}{2012}]{Kravtsov12}
{Kravtsov} A.~V.,  {Borgani} S.,  2012, \mn@doi [\araa]
  {10.1146/annurev-astro-081811-125502}, \href
  {http://adsabs.harvard.edu/abs/2012ARA%26A..50..353K} {50, 353}

\bibitem[\protect\citeauthoryear{{Lacey} \& {Cole}}{{Lacey} \&
  {Cole}}{1993}]{Lacey93}
{Lacey} C.,  {Cole} S.,  1993, \mnras, \href
  {http://adsabs.harvard.edu/abs/1993MNRAS.262..627L} {262, 627}

\bibitem[\protect\citeauthoryear{{Landry}, {Bonamente}, {Giles}, {Maughan},
  {Joy}  \& {Murray}}{{Landry} et~al.}{2013}]{Landry13}
{Landry} D.,  {Bonamente} M.,  {Giles} P.,  {Maughan} B.,  {Joy} M.,   {Murray}
  S.,  2013, \mn@doi [\mnras] {10.1093/mnras/stt901}, \href
  {https://ui.adsabs.harvard.edu/abs/2013MNRAS.433.2790L} {433, 2790}

\bibitem[\protect\citeauthoryear{{Liu} \& {Haiman}}{{Liu} \&
  {Haiman}}{2016}]{Liu16}
{Liu} J.,  {Haiman} Z.,  2016, \mn@doi [\prd] {10.1103/PhysRevD.94.043533},
  \href {http://adsabs.harvard.edu/abs/2016PhRvD..94d3533L} {94, 043533}

\bibitem[\protect\citeauthoryear{{Louren{\c{c}}o} et~al.,}{{Louren{\c{c}}o}
  et~al.}{2020}]{Lourenco20}
{Louren{\c{c}}o} A. C.~C.,  et~al., 2020, \mn@doi [\mnras]
  {10.1093/mnras/staa2464}, \href
  {https://ui.adsabs.harvard.edu/abs/2020MNRAS.498..835L} {498, 835}

\bibitem[\protect\citeauthoryear{{Mann} \& {Ebeling}}{{Mann} \&
  {Ebeling}}{2012}]{Mann12}
{Mann} A.~W.,  {Ebeling} H.,  2012, \mn@doi [\mnras]
  {10.1111/j.1365-2966.2011.20170.x}, \href
  {http://adsabs.harvard.edu/abs/2012MNRAS.420.2120M} {420, 2120}

\bibitem[\protect\citeauthoryear{{Markevitch} \& {Vikhlinin}}{{Markevitch} \&
  {Vikhlinin}}{2007}]{markevitch_viki07}
{Markevitch} M.,  {Vikhlinin} A.,  2007, \mn@doi [\physrep]
  {10.1016/j.physrep.2007.01.001}, \href
  {http://adsabs.harvard.edu/abs/2007PhR...443....1M} {443, 1}

\bibitem[\protect\citeauthoryear{{Markevitch}, {Gonzalez}, {Clowe},
  {Vikhlinin}, {Forman}, {Jones}, {Murray}  \& {Tucker}}{{Markevitch}
  et~al.}{2004}]{markevitch04}
{Markevitch} M.,  {Gonzalez} A.~H.,  {Clowe} D.,  {Vikhlinin} A.,  {Forman} W.,
   {Jones} C.,  {Murray} S.,   {Tucker} W.,  2004, \mn@doi [\apj]
  {10.1086/383178}, \href {http://adsabs.harvard.edu/abs/2004ApJ...606..819M}
  {606, 819}

\bibitem[\protect\citeauthoryear{{Marrone} et~al.,}{{Marrone}
  et~al.}{2012}]{Marrone12}
{Marrone} D.~P.,  et~al., 2012, \mn@doi [\apj] {10.1088/0004-637X/754/2/119},
  \href {https://ui.adsabs.harvard.edu/abs/2012ApJ...754..119M} {754, 119}

\bibitem[\protect\citeauthoryear{{Marshall}, {Hobson}, {Gull}  \&
  {Bridle}}{{Marshall} et~al.}{2002}]{LensEnt2}
{Marshall} P.~J.,  {Hobson} M.~P.,  {Gull} S.~F.,   {Bridle} S.~L.,  2002,
  \mn@doi [\mnras] {10.1046/j.1365-8711.2002.05685.x}, \href
  {http://adsabs.harvard.edu/abs/2002MNRAS.335.1037M} {335, 1037}

\bibitem[\protect\citeauthoryear{{Martel}, {Robichaud}  \& {Barai}}{{Martel}
  et~al.}{2014}]{martel14}
{Martel} H.,  {Robichaud} F.,   {Barai} P.,  2014, \mn@doi [\apj]
  {10.1088/0004-637X/786/2/79}, \href
  {http://adsabs.harvard.edu/abs/2014ApJ...786...79M} {786, 79}

\bibitem[\protect\citeauthoryear{Martin, Quinn  \& Park}{Martin
  et~al.}{2011}]{MCMCpack}
Martin A.~D.,  Quinn K.~M.,   Park J.~H.,  2011, Journal of Statistical
  Software, 42, 22

\bibitem[\protect\citeauthoryear{{Massey}, {Kitching}  \& {Nagai}}{{Massey}
  et~al.}{2011}]{Massey11}
{Massey} R.,  {Kitching} T.,   {Nagai} D.,  2011, \mn@doi [\mnras]
  {10.1111/j.1365-2966.2011.18246.x}, \href
  {https://ui.adsabs.harvard.edu/abs/2011MNRAS.413.1709M} {413, 1709}

\bibitem[\protect\citeauthoryear{{Maughan}, {Giles}, {Rines}, {Diaferio},
  {Geller}, {Van Der Pyl}  \& {Bonamente}}{{Maughan} et~al.}{2016}]{Maughan16}
{Maughan} B.~J.,  {Giles} P.~A.,  {Rines} K.~J.,  {Diaferio} A.,  {Geller}
  M.~J.,  {Van Der Pyl} N.,   {Bonamente} M.,  2016, \mn@doi [\mnras]
  {10.1093/mnras/stw1610}, \href
  {https://ui.adsabs.harvard.edu/abs/2016MNRAS.461.4182M} {461, 4182}

\bibitem[\protect\citeauthoryear{{Medezinski}, {Broadhurst}, {Umetsu}, {Oguri},
  {Rephaeli}  \& {Ben{\'{\i}}tez}}{{Medezinski} et~al.}{2010}]{med10}
{Medezinski} E.,  {Broadhurst} T.,  {Umetsu} K.,  {Oguri} M.,  {Rephaeli} Y.,
  {Ben{\'{\i}}tez} N.,  2010, \mn@doi [\mnras]
  {10.1111/j.1365-2966.2010.16491.x}, \href
  {http://adsabs.harvard.edu/abs/2010MNRAS.405..257M} {405, 257}

\bibitem[\protect\citeauthoryear{{Medezinski} et~al.,}{{Medezinski}
  et~al.}{2018}]{Medezinski18}
{Medezinski} E.,  et~al., 2018, \mn@doi [\pasj] {10.1093/pasj/psy009}, \href
  {http://adsabs.harvard.edu/abs/2018PASJ...70...30M} {70, 30}

\bibitem[\protect\citeauthoryear{{Mellier}}{{Mellier}}{1999}]{mellier99}
{Mellier} Y.,  1999, \mn@doi [\araa] {10.1146/annurev.astro.37.1.127}, \href
  {http://adsabs.harvard.edu/abs/1999ARA%26A..37..127M} {37, 127}

\bibitem[\protect\citeauthoryear{{Melnick} \& {Moles}}{{Melnick} \&
  {Moles}}{1987}]{Melnick87}
{Melnick} J.,  {Moles} M.,  1987, \rmxaa, \href
  {https://ui.adsabs.harvard.edu/abs/1987RMxAA..14...72M} {14, 72}

\bibitem[\protect\citeauthoryear{{Merten} et~al.,}{{Merten}
  et~al.}{2011}]{merten}
{Merten} J.,  et~al., 2011, \mn@doi [\mnras]
  {10.1111/j.1365-2966.2011.19266.x}, \href
  {http://adsabs.harvard.edu/abs/2011MNRAS.417..333M} {417, 333}

\bibitem[\protect\citeauthoryear{{Monteiro-Oliveira}, {Cypriano}, {Machado},
  {Lima Neto}, {Ribeiro}, {Sodr{\'e}}  \& {Dupke}}{{Monteiro-Oliveira}
  et~al.}{2017a}]{Monteiro-Oliveira17a}
{Monteiro-Oliveira} R.,  {Cypriano} E.~S.,  {Machado} R.~E.~G.,  {Lima Neto}
  G.~B.,  {Ribeiro} A.~L.~B.,  {Sodr{\'e}} L.,   {Dupke} R.,  2017a, \mn@doi
  [\mnras] {10.1093/mnras/stw3238}, \href
  {http://adsabs.harvard.edu/abs/2017MNRAS.466.2614M} {466, 2614}

\bibitem[\protect\citeauthoryear{{Monteiro-Oliveira}, {Lima Neto}, {Cypriano},
  {Machado}, {Capelato}, {Lagan{\'a}}, {Durret}  \&
  {Bagchi}}{{Monteiro-Oliveira} et~al.}{2017b}]{Monteiro-Oliveira17b}
{Monteiro-Oliveira} R.,  {Lima Neto} G.~B.,  {Cypriano} E.~S.,  {Machado}
  R.~E.~G.,  {Capelato} H.~V.,  {Lagan{\'a}} T.~F.,  {Durret} F.,   {Bagchi}
  J.,  2017b, \mn@doi [\mnras] {10.1093/mnras/stx791}, \href
  {http://adsabs.harvard.edu/abs/2017MNRAS.468.4566M} {468, 4566}

\bibitem[\protect\citeauthoryear{{Monteiro-Oliveira}, {Cypriano}, {Vitorelli},
  {Ribeiro}, {Sodr{\'e}}, {Dupke}  \& {Mendes de Oliveira}}{{Monteiro-Oliveira}
  et~al.}{2018}]{Monteiro-Oliveira18}
{Monteiro-Oliveira} R.,  {Cypriano} E.~S.,  {Vitorelli} A.~Z.,  {Ribeiro}
  A.~L.~B.,  {Sodr{\'e}} L.,  {Dupke} R.,   {Mendes de Oliveira} C.,  2018,
  \mn@doi [\mnras] {10.1093/mnras/sty2349}, \href
  {http://adsabs.harvard.edu/abs/2018MNRAS.481.1097M} {481, 1097}

\bibitem[\protect\citeauthoryear{{Monteiro-Oliveira}, {Doubrawa}, {Machado},
  {Lima Neto}, {Castejon}  \& {Cypriano}}{{Monteiro-Oliveira}
  et~al.}{2020}]{Monteiro-Oliveira20}
{Monteiro-Oliveira} R.,  {Doubrawa} L.,  {Machado} R.~E.~G.,  {Lima Neto}
  G.~B.,  {Castejon} M.,   {Cypriano} E.~S.,  2020, \mn@doi [\mnras]
  {10.1093/mnras/staa1218}, \href
  {https://ui.adsabs.harvard.edu/abs/2020MNRAS.495.2007M} {495, 2007}

\bibitem[\protect\citeauthoryear{{Moura}, {Machado}  \&
  {Monteiro-Oliveira}}{{Moura} et~al.}{2021}]{Moura20}
{Moura} M.~T.,  {Machado} R. E.~G.,   {Monteiro-Oliveira} R.,  2021, \mn@doi
  [\mnras] {10.1093/mnras/staa3399}, \href
  {https://ui.adsabs.harvard.edu/abs/2021MNRAS.500.1858M} {500, 1858}

\bibitem[\protect\citeauthoryear{{Munari}, {Biviano}, {Borgani}, {Murante}  \&
  {Fabjan}}{{Munari} et~al.}{2013}]{Munari13}
{Munari} E.,  {Biviano} A.,  {Borgani} S.,  {Murante} G.,   {Fabjan} D.,  2013,
  \mn@doi [\mnras] {10.1093/mnras/stt049}, \href
  {https://ui.adsabs.harvard.edu/abs/2013MNRAS.430.2638M} {430, 2638}

\bibitem[\protect\citeauthoryear{{Navarro}, {Frenk}  \& {White}}{{Navarro}
  et~al.}{1996}]{nfw96}
{Navarro} J.~F.,  {Frenk} C.~S.,   {White} S.~D.~M.,  1996, \mn@doi [\apj]
  {10.1086/177173}, \href {http://adsabs.harvard.edu/abs/1996ApJ...462..563N}
  {462, 563}

\bibitem[\protect\citeauthoryear{{Navarro}, {Frenk}  \& {White}}{{Navarro}
  et~al.}{1997}]{nfw97}
{Navarro} J.~F.,  {Frenk} C.~S.,   {White} S.~D.~M.,  1997, \apj, \href
  {http://adsabs.harvard.edu/abs/1997ApJ...490..493N} {490, 493}

\bibitem[\protect\citeauthoryear{{Nychka}, {Furrer}  \& {Sain}}{{Nychka}
  et~al.}{2014}]{fields}
{Nychka} D.,  {Furrer} R.,   {Sain} S.,  2014, Fields: Tools for spatial data.
\url {http://CRAN.R-project.org/package=fields}

\bibitem[\protect\citeauthoryear{{Okabe} \& {Smith}}{{Okabe} \&
  {Smith}}{2016}]{Okabe16}
{Okabe} N.,  {Smith} G.~P.,  2016, \mn@doi [\mnras] {10.1093/mnras/stw1539},
  \href {https://ui.adsabs.harvard.edu/abs/2016MNRAS.461.3794O} {461, 3794}

\bibitem[\protect\citeauthoryear{{Okabe}, {Takada}, {Umetsu}, {Futamase}  \&
  {Smith}}{{Okabe} et~al.}{2010}]{Okabe10}
{Okabe} N.,  {Takada} M.,  {Umetsu} K.,  {Futamase} T.,   {Smith} G.~P.,  2010,
  \mn@doi [\pasj] {10.1093/pasj/62.3.811}, \href
  {http://adsabs.harvard.edu/abs/2010PASJ...62..811O} {62, 811}

\bibitem[\protect\citeauthoryear{{Oke}}{{Oke}}{1974}]{Oke74}
{Oke} J.~B.,  1974, \mn@doi [\apjs] {10.1086/190287}, \href
  {https://ui.adsabs.harvard.edu/abs/1974ApJS...27...21O} {27, 21}

\bibitem[\protect\citeauthoryear{{Ouchi} et~al.,}{{Ouchi}
  et~al.}{2004}]{sdfred1}
{Ouchi} M.,  et~al., 2004, \mn@doi [\apj] {10.1086/422207}, \href
  {http://adsabs.harvard.edu/abs/2004ApJ...611..660O} {611, 660}

\bibitem[\protect\citeauthoryear{{Pandge}, {Monteiro-Oliveira}, {Bagchi},
  {Simionescu}, {Limousin}  \& {Raychaudhury}}{{Pandge}
  et~al.}{2019}]{Pandge19}
{Pandge} M.~B.,  {Monteiro-Oliveira} R.,  {Bagchi} J.,  {Simionescu} A.,
  {Limousin} M.,   {Raychaudhury} S.,  2019, \mn@doi [\mnras]
  {10.1093/mnras/sty2937}, \href
  {https://ui.adsabs.harvard.edu/abs/2019MNRAS.482.5093P} {482, 5093}

\bibitem[\protect\citeauthoryear{{Pinkney}, {Roettiger}, {Burns}  \&
  {Bird}}{{Pinkney} et~al.}{1996}]{pinkney}
{Pinkney} J.,  {Roettiger} K.,  {Burns} J.~O.,   {Bird} C.~M.,  1996, \mn@doi
  [\apjs] {10.1086/1972290}, \href
  {http://adsabs.harvard.edu/abs/1996ApJS..104....1P} {104, 1}

\bibitem[\protect\citeauthoryear{{Planck Collaboration} \& {Ade}}{{Planck
  Collaboration} \& {Ade}}{2013}]{Planck13}
{Planck Collaboration} {Ade} P.~A.~R. e.~a.,  2013, \mn@doi [\aap]
  {10.1051/0004-6361/201219398}, \href
  {http://adsabs.harvard.edu/abs/2013A%26A...550A.129P} {550, A129}

\bibitem[\protect\citeauthoryear{{Pratt}, {Arnaud}, {Biviano}, {Eckert},
  {Ettori}, {Nagai}, {Okabe}  \& {Reiprich}}{{Pratt} et~al.}{2019}]{Pratt19}
{Pratt} G.~W.,  {Arnaud} M.,  {Biviano} A.,  {Eckert} D.,  {Ettori} S.,
  {Nagai} D.,  {Okabe} N.,   {Reiprich} T.~H.,  2019, \mn@doi [\ssr]
  {10.1007/s11214-019-0591-0}, \href
  {https://ui.adsabs.harvard.edu/abs/2019SSRv..215...25P} {215, 25}

\bibitem[\protect\citeauthoryear{{Raychaudhury}}{{Raychaudhury}}{1989}]{Raychaudhury89}
{Raychaudhury} S.,  1989, \mn@doi [\nat] {10.1038/342251a0}, \href
  {https://ui.adsabs.harvard.edu/abs/1989Natur.342..251R} {342, 251}

\bibitem[\protect\citeauthoryear{{Reese} et~al.,}{{Reese}
  et~al.}{2012}]{Reese12}
{Reese} E.~D.,  et~al., 2012, \mn@doi [\apj] {10.1088/0004-637X/751/1/12},
  \href {https://ui.adsabs.harvard.edu/abs/2012ApJ...751...12R} {751, 12}

\bibitem[\protect\citeauthoryear{{Ribeiro}, {de Carvalho}, {Trevisan},
  {Capelato}, {La Barbera}, {Lopes}  \& {Schilling}}{{Ribeiro}
  et~al.}{2013}]{Ribeiro13}
{Ribeiro} A.~L.~B.,  {de Carvalho} R.~R.,  {Trevisan} M.,  {Capelato} H.~V.,
  {La Barbera} F.,  {Lopes} P.~A.~A.,   {Schilling} A.~C.,  2013, \mn@doi
  [\mnras] {10.1093/mnras/stt1071}, \href
  {https://ui.adsabs.harvard.edu/\#abs/2013MNRAS.434..784R} {434, 784}

\bibitem[\protect\citeauthoryear{{Rines}, {Geller}, {Diaferio}  \&
  {Kurtz}}{{Rines} et~al.}{2013}]{Rines13}
{Rines} K.,  {Geller} M.~J.,  {Diaferio} A.,   {Kurtz} M.~J.,  2013, \mn@doi
  [\apj] {10.1088/0004-637X/767/1/15}, \href
  {https://ui.adsabs.harvard.edu/#abs/2013ApJ...767...15R} {767, 15}

\bibitem[\protect\citeauthoryear{{Sagunski}, {Gad-Nasr}, {Colquhoun},
  {Robertson}  \& {Tulin}}{{Sagunski} et~al.}{2020}]{Sagunski20}
{Sagunski} L.,  {Gad-Nasr} S.,  {Colquhoun} B.,  {Robertson} A.,   {Tulin} S.,
  2020, arXiv e-prints, \href
  {https://ui.adsabs.harvard.edu/abs/2020arXiv200612515S} {p. arXiv:2006.12515}

\bibitem[\protect\citeauthoryear{{Scaramella}, {Baiesi-Pillastrini},
  {Chincarini}, {Vettolani}  \& {Zamorani}}{{Scaramella}
  et~al.}{1989}]{Scaramella89}
{Scaramella} R.,  {Baiesi-Pillastrini} G.,  {Chincarini} G.,  {Vettolani} G.,
  {Zamorani} G.,  1989, \mn@doi [\nat] {10.1038/338562a0}, \href
  {https://ui.adsabs.harvard.edu/abs/1989Natur.338..562S} {338, 562}

\bibitem[\protect\citeauthoryear{Scrucca, Fop, Murphy  \& Raftery}{Scrucca
  et~al.}{2016}]{mclust}
Scrucca L.,  Fop M.,  Murphy T.~B.,   Raftery A.~E.,  2016, \mn@doi [{The R
  Journal}] {10.32614/RJ-2016-021}, 8, 289

\bibitem[\protect\citeauthoryear{{Seitz}, {Schneider}  \& {Bartelmann}}{{Seitz}
  et~al.}{1998}]{seitz98}
{Seitz} S.,  {Schneider} P.,   {Bartelmann} M.,  1998, \aap, \href
  {http://adsabs.harvard.edu/abs/1998A%26A...337..325S} {337, 325}

\bibitem[\protect\citeauthoryear{{Sheth}, {Mo}  \& {Tormen}}{{Sheth}
  et~al.}{2001}]{Sheth01}
{Sheth} R.~K.,  {Mo} H.~J.,   {Tormen} G.,  2001, \mn@doi [\mnras]
  {10.1046/j.1365-8711.2001.04006.x}, \href
  {https://ui.adsabs.harvard.edu/abs/2001MNRAS.323....1S} {323, 1}

\bibitem[\protect\citeauthoryear{{Sif{\'o}n} et~al.,}{{Sif{\'o}n}
  et~al.}{2016}]{Sifon16}
{Sif{\'o}n} C.,  et~al., 2016, \mn@doi [\mnras] {10.1093/mnras/stw1284}, \href
  {https://ui.adsabs.harvard.edu/abs/2016MNRAS.461..248S} {461, 248}

\bibitem[\protect\citeauthoryear{{Soja}, {Sodr{\'e}}, {Monteiro-Oliveira},
  {Cypriano}  \& {Lima Neto}}{{Soja} et~al.}{2018}]{Soja18}
{Soja} A.~C.,  {Sodr{\'e}} L.,  {Monteiro-Oliveira} R.,  {Cypriano} E.~S.,
  {Lima Neto} G.~B.,  2018, \mn@doi [\mnras] {10.1093/mnras/sty638}, \href
  {http://adsabs.harvard.edu/abs/2018MNRAS.477.3279S} {477, 3279}

\bibitem[\protect\citeauthoryear{{Springel} et~al.,}{{Springel}
  et~al.}{2018}]{Springel18}
{Springel} V.,  et~al., 2018, \mn@doi [\mnras] {10.1093/mnras/stx3304}, \href
  {http://adsabs.harvard.edu/abs/2018MNRAS.475..676S} {475, 676}

\bibitem[\protect\citeauthoryear{{Swarup}, {Ananthakrishnan}, {Kapahi}, {Rao},
  {Subrahmanya}  \& {Kulkarni}}{{Swarup} et~al.}{1991}]{GMRT}
{Swarup} G.,  {Ananthakrishnan} S.,  {Kapahi} V.~K.,  {Rao} A.~P.,
  {Subrahmanya} C.~R.,   {Kulkarni} V.~K.,  1991, Current Science, \href
  {https://ui.adsabs.harvard.edu/abs/1991CuSc...60...95S} {60, 95}

\bibitem[\protect\citeauthoryear{{Takizawa}, {Nagino}  \&
  {Matsushita}}{{Takizawa} et~al.}{2010}]{Takizawa10}
{Takizawa} M.,  {Nagino} R.,   {Matsushita} K.,  2010, \mn@doi [\pasj]
  {10.1093/pasj/62.4.951}, \href
  {http://adsabs.harvard.edu/abs/2010PASJ...62..951T} {62, 951}

\bibitem[\protect\citeauthoryear{{Umetsu}, {Zitrin}, {Gruen}, {Merten},
  {Donahue}  \& {Postman}}{{Umetsu} et~al.}{2016}]{Umetsu16}
{Umetsu} K.,  {Zitrin} A.,  {Gruen} D.,  {Merten} J.,  {Donahue} M.,
  {Postman} M.,  2016, \mn@doi [\apj] {10.3847/0004-637X/821/2/116}, \href
  {https://ui.adsabs.harvard.edu/abs/2016ApJ...821..116U} {821, 116}

\bibitem[\protect\citeauthoryear{{Venturi}, {Giacintucci}, {Brunetti},
  {Cassano}, {Bardelli}, {Dallacasa}  \& {Setti}}{{Venturi}
  et~al.}{2007}]{Venturi07}
{Venturi} T.,  {Giacintucci} S.,  {Brunetti} G.,  {Cassano} R.,  {Bardelli} S.,
   {Dallacasa} D.,   {Setti} G.,  2007, \mn@doi [\aap]
  {10.1051/0004-6361:20065961}, \href
  {https://ui.adsabs.harvard.edu/abs/2007A&A...463..937V} {463, 937}

\bibitem[\protect\citeauthoryear{{Wei}, {Li}, {Kang}, {Liu}, {Fan}, {Yuan}  \&
  {Pan}}{{Wei} et~al.}{2018}]{Wei18}
{Wei} C.,  {Li} G.,  {Kang} X.,  {Liu} X.,  {Fan} Z.,  {Yuan} S.,   {Pan} C.,
  2018, \mn@doi [\mnras] {10.1093/mnras/sty1268}, \href
  {http://adsabs.harvard.edu/abs/2018MNRAS.478.2987W} {478, 2987}

\bibitem[\protect\citeauthoryear{{Wen} \& {Han}}{{Wen} \& {Han}}{2013}]{Wen13}
{Wen} Z.~L.,  {Han} J.~L.,  2013, \mn@doi [\mnras] {10.1093/mnras/stt1581},
  \href {https://ui.adsabs.harvard.edu/abs/2013MNRAS.436..275W} {436, 275}

\bibitem[\protect\citeauthoryear{{Wittman}, {Cornell}  \& {Nguyen}}{{Wittman}
  et~al.}{2018}]{Wittman18a}
{Wittman} D.,  {Cornell} B.~H.,   {Nguyen} J.,  2018, \mn@doi [\apj]
  {10.3847/1538-4357/aacf3e}, \href
  {http://adsabs.harvard.edu/abs/2018ApJ...862..160W} {862, 160}

\bibitem[\protect\citeauthoryear{{Wojtak}, {{\L}okas}, {Mamon},
  {Gottl{\"o}ber}, {Prada}  \& {Moles}}{{Wojtak} et~al.}{2007}]{Wojtak07}
{Wojtak} R.,  {{\L}okas} E.~L.,  {Mamon} G.~A.,  {Gottl{\"o}ber} S.,  {Prada}
  F.,   {Moles} M.,  2007, \mn@doi [\aap] {10.1051/0004-6361:20066813}, \href
  {http://adsabs.harvard.edu/abs/2007A%26A...466..437W} {466, 437}

\bibitem[\protect\citeauthoryear{{Wright}}{{Wright}}{2006}]{CosmoCalc}
{Wright} E.~L.,  2006, \mn@doi [\pasp] {10.1086/510102}, \href
  {http://adsabs.harvard.edu/abs/2006PASP..118.1711W} {118, 1711}

\bibitem[\protect\citeauthoryear{{Wright} \& {Brainerd}}{{Wright} \&
  {Brainerd}}{2000}]{wright00}
{Wright} C.~O.,  {Brainerd} T.~G.,  2000, \mn@doi [\apj] {10.1086/308744},
  \href {http://adsabs.harvard.edu/abs/2000ApJ...534...34W} {534, 34}

\bibitem[\protect\citeauthoryear{{Yahil} \& {Vidal}}{{Yahil} \&
  {Vidal}}{1977}]{3sigmaclip}
{Yahil} A.,  {Vidal} N.~V.,  1977, \mn@doi [\apj] {10.1086/155257}, \href
  {http://adsabs.harvard.edu/abs/1977ApJ...214..347Y} {214, 347}

\bibitem[\protect\citeauthoryear{{Yang}, {Kratochvil}, {Huffenberger}, {Haiman}
   \& {May}}{{Yang} et~al.}{2013}]{Yang13}
{Yang} X.,  {Kratochvil} J.~M.,  {Huffenberger} K.,  {Haiman} Z.,   {May} M.,
  2013, \mn@doi [\prd] {10.1103/PhysRevD.87.023511}, \href
  {https://ui.adsabs.harvard.edu/abs/2013PhRvD..87b3511Y} {87, 023511}

\bibitem[\protect\citeauthoryear{{York} et~al.,}{{York} et~al.}{2000}]{York00}
{York} D.~G.,  et~al., 2000, \mn@doi [\aj] {10.1086/301513}, \href
  {https://ui.adsabs.harvard.edu/abs/2000AJ....120.1579Y} {120, 1579}

\bibitem[\protect\citeauthoryear{{Zacharias}, {Finch}, {Girard}, {Henden},
  {Bartlett}, {Monet}  \& {Zacharias}}{{Zacharias} et~al.}{2013}]{Zacharias13}
{Zacharias} N.,  {Finch} C.~T.,  {Girard} T.~M.,  {Henden} A.,  {Bartlett}
  J.~L.,  {Monet} D.~G.,   {Zacharias} M.~I.,  2013, \mn@doi [\aj]
  {10.1088/0004-6256/145/2/44}, \href
  {https://ui.adsabs.harvard.edu/abs/2013AJ....145...44Z} {145, 44}

\bibitem[\protect\citeauthoryear{{Zhang}, {B{\"o}hringer}, {Finoguenov},
  {Ikebe}, {Matsushita}, {Schuecker}, {Guzzo}  \& {Collins}}{{Zhang}
  et~al.}{2006}]{Zhang06}
{Zhang} Y.~Y.,  {B{\"o}hringer} H.,  {Finoguenov} A.,  {Ikebe} Y.,
  {Matsushita} K.,  {Schuecker} P.,  {Guzzo} L.,   {Collins} C.~A.,  2006,
  \mn@doi [\aap] {10.1051/0004-6361:20053650}, \href
  {https://ui.adsabs.harvard.edu/abs/2006A&A...456...55Z} {456, 55}

\bibitem[\protect\citeauthoryear{{Zhang}, {Finoguenov}, {B{\"o}hringer},
  {Kneib}, {Smith}, {Kneissl}, {Okabe}  \& {Dahle}}{{Zhang}
  et~al.}{2008}]{Zhang08}
{Zhang} Y.~Y.,  {Finoguenov} A.,  {B{\"o}hringer} H.,  {Kneib} J.~P.,  {Smith}
  G.~P.,  {Kneissl} R.,  {Okabe} N.,   {Dahle} H.,  2008, \mn@doi [\aap]
  {10.1051/0004-6361:20079103}, \href
  {https://ui.adsabs.harvard.edu/abs/2008A&A...482..451Z} {482, 451}

\bibitem[\protect\citeauthoryear{{von der Linden} et~al.,}{{von der Linden}
  et~al.}{2014}]{vonderLinden14}
{von der Linden} A.,  et~al., 2014, \mn@doi [\mnras] {10.1093/mnras/stt1945},
  \href {https://ui.adsabs.harvard.edu/abs/2014MNRAS.439....2V} {439, 2}

\makeatother
\end{thebibliography}




\bsp	
\label{lastpage}
\end{document}